\documentclass[aps,prx,reprint,superscriptaddress,longbibliography,
               nofootinbib]{revtex4-2}

\usepackage{amsmath,amssymb}
\usepackage{bm}
\usepackage{graphicx}
\usepackage{booktabs}
\usepackage{algpseudocode}
\makeatletter
\newcounter{algorithm}
\renewcommand{\thealgorithm}{\arabic{algorithm}}
\newenvironment{algorithm}[1][t]{
  \begin{figure}[#1]\footnotesize
  \refstepcounter{algorithm}
  \hrule height.8pt \vspace{3pt}
  \def\caption##1{\noindent\textbf{Algorithm~\thealgorithm:} ##1\par
    \vspace{3pt}\hrule height.4pt \vspace{2pt}}
}{\vspace{2pt}\hrule height.8pt \end{figure}}
\makeatother
\providecommand{\Return}{\textbf{return}~}
\usepackage[colorlinks=true,linkcolor=blue,citecolor=blue,urlcolor=blue]{hyperref}
\usepackage{xcolor}

\usepackage{tikz}
\usetikzlibrary{arrows.meta,positioning,calc,bending}

\newcommand{\bmu}{\boldsymbol{\mu}}

\newcommand{\bfx}{\mathbf{x}}
\newcommand{\bfy}{\mathbf{y}}
\newcommand{\bfzero}{\mathbf{0}}
\newcommand{\bfI}{\mathbf{I}}

\newcommand{\bfz}{\mathbf{z}}
\newcommand{\xhat}{\hat{\mathbf{x}}_0}

\newcommand{\epshat}{\hat{\epsilon}}
\newcommand{\net}{\epsilon_\theta}
\newcommand{\ddpmstep}{\operatorname{DDPM-STEP}}
\newcommand{\ET}{E_T}

\begin{document}

\title{Statistical validation of calorimeter inpainting with generative diffusion priors }

\author{Himanshu Raj}
\author{Roli Esha}
\affiliation{Department of Physics and Astronomy, Stony Brook University, Stony Brook, NY 11794, USA}
\affiliation{Center for Frontiers in Nuclear Science, Stony Brook University, Stony Brook, NY 11794, USA}

\begin{abstract}
Localized detector inefficiencies produce incomplete calorimeter data that limit the ability to perform precision measurements. We address this problem in relativistic heavy-ion collisions from a Bayesian perspective using pretrained calorimeter diffusion models as priors to reconstruct the missing signal conditioned on surrounding measurements. In this work, we conduct a systematic comparison of several diffusion-based inpainting algorithms, whose performance is evaluated using Bayesian posterior diagnostics of energy response, spatial bias, and uncertainty calibration. The reconstruction fidelity is also analyzed across collision centralities and masked region sizes. This study establishes a general validation strategy for probabilistic reconstruction of missing detector information. 
\end{abstract}

\maketitle

\section{Introduction}
\label{sec:intro}

Many scientific inference problems require recovering information that cannot be directly observed. In modern high-energy and nuclear physics experiments, this can arise from missing detector channels or incomplete detector coverage. These inverse problems are inherently ill-posed because multiple solutions can be consistent with the available measurements. The goal is therefore not only to find a plausible reconstruction, but also to characterize the range of solutions consistent with both the observations and prior knowledge.

Recent advances in deep generative models have provided a principled framework for solving such problems. In particular, score-based diffusion models~\cite{vincent2011connection, sohl2015deep, ho2020ddpm, song2021sde} learn complex, high-dimensional data distributions and generate ensembles of statistically consistent reconstructions rather than a single deterministic estimate~\cite{mikuni2022caloscore,amram2023calodiffusion,torbunov2024calo}\footnote{See also~\cite{paganini2018calogan,krause2021caloflow} for other deep generative approaches that include Generative Adversarial Networks and Normalizing Flows.}. These models have the ability to approximate the posterior distribution of unknown missing signal, making them especially attractive for scientific applications where uncertainty quantification is as important as reconstruction accuracy. Several training-free diffusion-based `inpainting' algorithms have been developed to solve such problems~\cite{lugmayr2022repaint,kawar2022ddrm,wang2022ddnm,chung2022mcg,chung2023dps,song2023pigdm}. Because these methods operate directly on a pretrained diffusion model, the same generative prior can be reused for different detector configurations and masking geometries.

Calorimeter reconstruction in high-energy nuclear physics provides an ideal testbed for these methods. Inactive or dead regions in calorimeters lead to missing energy deposits that one may need to reconstruct before precision measurements can be performed. For instance, since calorimeter energies directly enter jet reconstruction, electromagnetic probes, and global event observables, reconstruction uncertainties can propagate to many downstream physics analyses. At the Relativistic Heavy Ion Collider (RHIC) and the Large Hadron Collider (LHC), inactive calorimeter cells are a well-known source of systematic uncertainty in jet energy calibration and missing transverse momentum reconstruction~\cite{atlas2018jes,cms2019met}. In heavy-ion collisions, where energetic jets are embedded in a large fluctuating background, missing detector regions can additionally bias jet spectra, jet substructure, and event-by-event observables~\cite{busza2018heavy,connors2018jetreview}.

Traditional approaches typically rely on fiducial cuts, event rejection, or interpolation from neighboring calorimeter cells~\cite{atlas2018jes,cms2019met,connors2018jetreview,Boldyrev:2024neh}. While simple to implement, these methods either reduce the available statistics or provide only a single deterministic estimate of the missing detector response. As a result, they provide no direct way to quantify or propagate the uncertainty of the reconstruction.

Unlike many image restoration problems in computer vision, calorimeter inpainting is not primarily a denoising problem. The measurements in active calorimeter towers are treated as known, while only the energies in inactive regions are unknown. The task is therefore to infer physically plausible values for the missing detector response that remain consistent with both the measured data and the underlying event topology. This motivates a Bayesian formulation in which the goal is to infer the distribution of the missing detector response rather than a single completed image.

Despite rapid progress in diffusion-based reconstruction, most existing evaluations rely almost exclusively on deterministic image-quality metrics such as mean squared error, peak signal-to-noise ratio, structural similarity~\cite{wang2004image}, Learned Perceptual Image Patch Similarity \cite{abs-1801-03924} or Frechet Inception Distance \cite{HeuselRUNKH17}. While these metrics measure reconstruction fidelity, they do not determine whether the predicted distribution is statistically calibrated. Two algorithms may therefore achieve similar reconstruction accuracy while providing substantially different uncertainty estimates, leading to different conclusions when the results are propagated to downstream physics analyses.

Simulation-Based Calibration (SBC) provides a rigorous statistical framework for addressing this problem~\cite{talts2018sbc}. When the ground-truth data are generated from the same prior used for posterior inference, exact posterior samples are statistically exchangeable with the truth, and departures from calibration reflect approximation errors in the inference algorithm. Diagnostics such as standardized residuals ($z$-scores), the Probability Integral Transform (PIT), empirical coverage, posterior sharpness, and posterior shrinkage therefore provide complementary tests of reconstruction bias, uncertainty calibration, and predictive reliability~\cite{gelman2013bda}. Although widely used in Bayesian statistics, these tools have seen limited application in evaluating generative models for scientific inverse problems.

In this work, we present a comprehensive statistical validation of diffusion-based calorimeter inpainting within the SBC framework. Using calorimeter images generated by a DDPM trained on an sPHENIX-inspired electromagnetic calorimeter for Au+Au collisions at $\sqrt{s_{NN}}=200$~GeV~\cite{torbunov2024calo}, we systematically compare various state-of-the-art training-free diffusion inpainting algorithms under identical reconstruction conditions. Their performance is assessed using complementary deterministic and probabilistic diagnostics that quantify both reconstruction accuracy and uncertainty calibration. We further examine the robustness of one of the best-performing methods across collision centralities and masked detector region sizes ranging from $6\times6$ to $12\times12$ towers. This approach provides a quantitative way to determine whether an inpainting model reproduces the missing energy and assigns reliable uncertainties, while providing a general framework for validating probabilistic reconstruction methods in nuclear and high-energy physics.

The remainder of this paper is organized as follows. Section~\ref{sec:setup} introduces the denoising diffusion probabilistic model and the calorimeter dataset used in this study. Section~\ref{sec:algorithms} presents the inpainting algorithms employed. Section~\ref{sec:results} discusses the reconstruction results, including comparisons of the different inpainting algorithms, studies of centrality dependence, and the dependence on the size of the dead detector region. Finally, Section~\ref{sec:conclusion} summarizes the conclusions and discusses future directions.

\section{Diffusion model}
\label{sec:setup}

\subsection{The Denoising Diffusion Probabilistic Model}
\label{sec:DDPM-setup}

\begin{figure}[ht]
\centering
\begin{tikzpicture}[>=Latex, image/.style={inner sep=0pt}, label/.style={font=\scriptsize}]

\node[image] (x0)
    {\includegraphics[width=1.1cm]{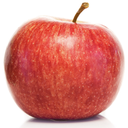}};

\node[image, right=0.10cm of x0] (x1)
    {\includegraphics[width=1.1cm]{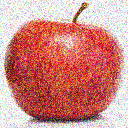}};

\node[right=0.12cm of x1, inner sep=0pt] (dots1)
    {$\cdots$};

\node[image, right=0.12cm of dots1] (xtm1)
    {\includegraphics[width=1.1cm]{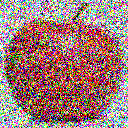}};

\node[image, right=0.10cm of xtm1] (xt)
    {\includegraphics[width=1.1cm]{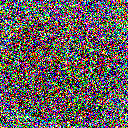}};

\node[right=0.12cm of xt, inner sep=0pt] (dots2)
    {$\cdots$};

\node[image, right=0.12cm of dots2] (xT)
    {\includegraphics[width=1.1cm]{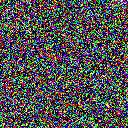}};

% State labels
\node[label, above=0.6mm of x0]   {$\bfx_0$};
\node[label, above=0.6mm of x1]   {$\bfx_1$};
\node[label, above=0.6mm of xtm1] {$\bfx_{t-1}$};
\node[label, above=0.6mm of xt]   {$\bfx_t$};
\node[label, above=0.6mm of xT]   {$\bfx_T$};

% One-step forward transition
\draw[->]
    (xtm1.north)
    to[bend left=40]
    node[above=2mm, font=\scriptsize]
    {$q(\bfx_t\mid \bfx_{t-1})$}
    (xt.north);

% One-step reverse transition
\draw[->]
    (xt.south)
    to[bend left=40]
    node[below, font=\scriptsize]
    {$p_\theta(\bfx_{t-1}\mid \bfx_t)$}
    (xtm1.south);
    
\end{tikzpicture}
\caption{Forward noising and the reverse denoising process in a diffusion model. The arrows show the fixed one-step forward transition \(q(\bfx_t\mid \bfx_{t-1})\) and the learned reverse transition \(p_\theta(\bfx_{t-1}\mid \bfx_t)\).}
\label{fig:diffusion_process}
\end{figure}
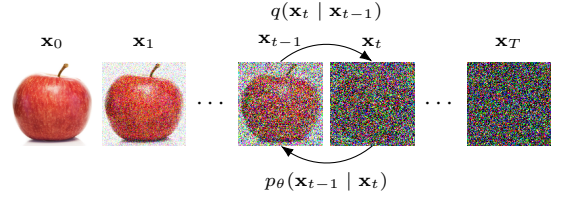

Diffusion models \cite{vincent2011connection, sohl2015deep, ho2020ddpm, song2021sde} are a family of deep generative models that are defined by a Markov chain structure. These models are trained by sequentially adding Gaussian noise to data samples via a fixed but tunable diffusion schedule. The model learns to reverse the Markov chain via a denoising process which is optimized by regressing the model's predicted noise against the added Gaussian noise. Figure \ref{fig:diffusion_process} depicts this process schematically. The learned model approximates the underlying data distribution from which a new data point can be sampled. These models have seen tremendous success primarily in image \cite{dhariwal2021diffusion, meng2021sdedit, rombach2022high} and video generation \cite{ho2022video, ho2022imagen} but also in other modalities that include audio synthesis (speech \cite{kong2020diffwave}, music \cite{mariani2023multi, evans2024long}) and other sequential data (text \cite{li2022diffusion}, time series \cite{tashiro2021csdi}). Our current work is based on Denoising Diffusion Probabilistic Models (DDPMs) \cite{ho2020ddpm, nichol2021improved}, a particular class of diffusion models, that we briefly summarize in this Section.

Given a sample $\mathbf{x}_0$ from some data distribution $q(\mathbf{x})$ we define a set of associated latent random variables $ \mathbf{x}_1, \mathbf{x}_2, \cdots , \mathbf{x}_T$ that are drawn from the following (forward) conditional distribution in the variance-preserving parameterization\footnote{Some papers use the equivalent variance-exploding parameterization where the forward conditionals take the form $q(\mathbf{x}_t | \mathbf{x}_{0})  = \mathcal{N}(\mathbf{x}_t; \mathbf{x}_{0}, \sigma_t^2 \mathbf{I})$ \cite{kawar2022ddrm}. The change of variables is reviewed in Appendix \ref{app:diffusion:vpve}.} 
\begin{align}
\label{eq:fwd_MC}
    q(\mathbf{x}_t | \mathbf{x}_{t-1})  = \mathcal{N}(\mathbf{x}_t; \sqrt{\alpha_{t}}\mathbf{x}_{t-1}, (1-\alpha_{t}) \mathbf{I})~.
\end{align}
This leads to a one-step forward process
\begin{align}
\label{eq:fwd_process}
    \mathbf{x}_t =  \sqrt{\alpha_{t}}\mathbf{x}_{t-1} + \sqrt{1-\alpha_{t}} ~\epsilon_t~,\quad \epsilon_t \sim \mathcal{N}(\bfzero, \bfI)~.
\end{align}
This is a `time' dependent noising process that starts from a data sample at step $0$ and undergoes successive Gaussian noise addition (with mean $\sqrt{\alpha_{t}}\mathbf{x}_{t-1}$ and variance $(1-\alpha_{t}) \mathbf{I}$ at step $t$) and ends in a heavily noisy version of it at time $T$ which is a hyperparameter. In the original DDPM formulation, the hyper-parameters $\alpha_t$ (usually expressed as $\beta_t := 1-\alpha_t$) that control the scale and variance of the noising schedule at timestep $t$ is taken to be linearly increasing in $\beta_t$: $\beta_1 < \beta_2 < \cdots < \beta_T$ (though other schedules have also been studied \cite{nichol2021improved}). A noised sample $\mathbf{x}_t$ at step $t$ can be written in closed form 
\begin{align}
\label{eq:x_t_fast_fwd}
    \mathbf{x}_t (\mathbf{x}_{0})  = \sqrt{\bar\alpha_{t}}\mathbf{x}_{0} + \sqrt{1-\bar \alpha_{t}}~\epsilon~,\qquad \epsilon \sim \mathcal{N}(\mathbf{0}, \mathbf{I})~,
\end{align}
where $\bar{\alpha}_t = \prod_{i=1}^t \alpha_i$. This follows from the properties of the Gaussian distribution. 

The goal is to learn to reverse this noising process. The reverse conditional $q(\mathbf{x}_{t-1} | \mathbf{x}_{t})$ obtained via Bayes' rule
\begin{align}
\label{eq:true_posterior}
    q(\mathbf{x}_{t-1} | \mathbf{x}_{t})  = \frac{q(\mathbf{x}_t | \mathbf{x}_{t-1}) q(\mathbf{x}_{t-1})}{q(\mathbf{x}_t)}
\end{align}
is not tractable since, at step $t$, one needs information of all possible $\mathbf{x}_{t-1}$ in order to obtain the denominator $q(\mathbf{x}_t)$: $q(\mathbf{x}_t)=\int d\mathbf{x}_{t-1}q(\mathbf{x}_t | \mathbf{x}_{t-1}) q(\mathbf{x}_{t-1}) $. Therefore a probabilistic denoising model $p_\theta(\mathbf{x}_{t-1} | \mathbf{x}_t)$ is introduced
\begin{align}
\label{eq:variational_posterior}
p_\theta(\mathbf{x}_{t-1} | \mathbf{x}_t) = \mathcal{N}(\mathbf{x}_{t-1}; \mu_\theta(\mathbf{x}_t, t), \Sigma_\theta(\mathbf{x}_t, t))~.
\end{align}
This again has the structure of a Markov chain which now runs in the reverse direction. The learnable parameters of the normal distribution $\mu_\theta(\mathbf{x}_t, t)$ and $\Sigma_\theta(\mathbf{x}_t, t)$ are parametrized by a neural network\footnote{Note that we do not have a subscript $t$ in $\theta$ because in practice the neural network parameters are amortized over the full range of $t$ (the weights of a single network are shared across all time steps).}. Usually, instead of parameterizing the reverse process to estimate the mean $\mu_\theta\left(\boldsymbol{x}_t, t\right)$, it is common to reparameterize it to predict the noise $\epsilon$ at step $t$ that was added to $\mathbf{x}_0$ to produce $\mathbf{x}_t$ in Eq.~\eqref{eq:x_t_fast_fwd}. Given a noisy sample $\bfx_t$ at step $t$ and denoting the denoising model by $\epsilon_\theta\left(\mathbf{x}_t, t\right)$, the mean  $\mu_\theta\left(\mathbf{x}_t, t\right)$ of the denoised sample at step $t-1$ takes the following form (see Appendix \ref{app:diffusion:posterior} for details)
\begin{align}
\label{eq:mu_reparam}
\mu_\theta\left(\mathbf{x}_t,t\right)=\frac{1}{\sqrt{\alpha_t}}\left(\mathbf{x}_t-\frac{1-\alpha_t}{\sqrt{1-\bar{\alpha}_t}} \epsilon_\theta\left(\mathbf{x}_t, t\right)\right)~.
\end{align}
A further design choice is to set the variance of the reverse process at time $t$ equal to\footnote{Another choice is to set the reverse variance equal to the forward variance $\Sigma_\theta\left(\mathbf{x}_t\right) = \beta_t \bfI $. This choice was implemented in the work of \cite{torbunov2024calo}. In practice there is little difference between the two choices.}
\begin{align}\Sigma_\theta\left(\mathbf{x}_t\right):=\tilde \beta_t \mathbf{I}~,\qquad \tilde \beta_t = \frac{1-\bar{\alpha}_{t-1}}{1-\bar{\alpha}_t} \beta_t
\end{align}
This is a principled choice that comes from the conditional $q(\bfx_{t-1}\mid \bfx_t, \bfx_0)$ which can be expressed as Gaussian by the Bayes' rule (see Appendix \ref{app:diffusion:posterior}).

The DDPM objective function $\ell(\theta)$ used for training the denoising model takes the form of a mean squared error between the added noise $\epsilon$ and the predicted noise $\epsilon_\theta\left(\mathbf{x}_t, t\right)$
\begin{align}
\label{eq:DDPM_objective}
\ell(\theta)=\underset{t, \epsilon, \mathbf{x}_0}{\mathbb{E}}\left[\left\|\epsilon-\epsilon_\theta\left(\mathbf{x}_t, t\right)\right\|_2^2\right] ~,
\end{align}
with the expectation taken over all time steps $t$, added noise $\epsilon$ and data samples $\mathbf{x}_0$. %In the above formulas $\mathbf{x}_t$ is given by \eqref{eq:x_t_fast_fwd}. 

Once the denoising model $p_\theta(\mathbf{x}_{t-1} | \mathbf{x}_t)$ is trained, a new sample can be generated by first sampling $\tilde{\mathbf{x}}_T \sim \mathcal{N}\left(0, \mathbf{I}\right)$ and then auto-regressively sampling from $\mathcal{N}\left(\mathbf{x}_{t-1} ; \mu_\theta\left(\tilde{\mathbf{x}}_t\right), \tilde \beta_t \mathbf{I}\right)$ to get the clean sample $\tilde{\mathbf{x}}_0$. We have collected further details in Appendix \ref{app:diffusion}.

\subsection{The Calorimeter DDPM}
\label{sec:calo-ddpm-review}

\begin{figure*}[ht]
    \centering
    \includegraphics[width=0.5\textwidth]{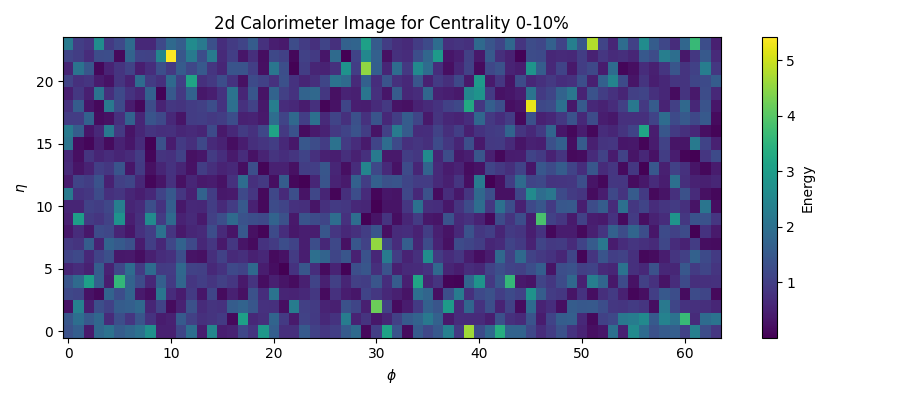}\hfill
    \includegraphics[width=0.5\textwidth]{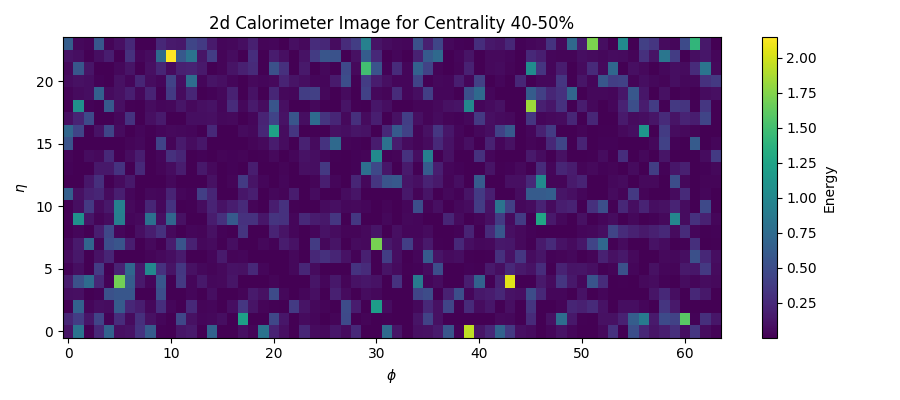}
    \caption{Calorimeter events generated from the trained DDPMs of \cite{torbunov2024calo} with $S=T=8000$ reverse steps. The image on the left is for centrality $0-10\%$ and the right is for centrality $40-50\%$.}
    \label{fig:event_display}
\end{figure*}

Our work builds upon the pre-trained whole-event calorimeter DDPMs (calo-DDPM) of \cite{torbunov2024calo}. This paper was the first to introduce the use of Diffusion model (along with Generative Adversarial Network which will not be the subject of our paper) for full-detector, whole event heavy-ion simulation trained on the simulated calorimeter data for the sPHENIX experiment at RHIC. In this subsection we briefly describe their work reviewing how their reference training data was obtained, the neural network configuration for training and inference that will be used later in Section \ref{sec:algorithms} for the inpainting study. Establishing this is essential because the true events we later inpaint are themselves drawn from these same networks - every property of the reference training dataset and every choice made during their training propagates directly into the problem of conditional in-painting.

\subsubsection{Dataset}
\label{sec:setup:sec1}

The upstream training dataset is obtained from a simulation of Au$+$Au collisions at a center-of-mass energy of $\sqrt{s_{NN}} = 200$~GeV~\cite{busza2018heavy}. These are produced via the \textsc{Hijing} Monte Carlo event generator~\cite{PhysRevD.44.3501}. The resulting long-lived particles are then propagated through a \textsc{Geant4}~\cite{{AGOSTINELLI2003250,1610988, ALLISON2016186}} model of the sPHENIX detector, so that detector response is folded into the truth at the level of energy deposited in the calorimeters.  Events are generated across a range of centralities, which in \textsc{Hijing} is set by the percentile of the impact parameter $b$ between the colliding nuclei and which controls the size of the produced medium and hence the overall event activity. Two centrality windows were considered: the $0-10\%$ ``central'' collision ($b = 0-4.88$~fm) that corresponds to highest multiplicity collisions, and the $40-50\%$ ``mid-central'' collision ($b = 9.71-10.81$~fm), with a correspondingly smaller event activity. Multiple simultaneous collisions within a beam bunch (``pileup'') were allowed, so the samples reflect realistic in-time event overlap. Because the two centrality classes have markedly different event characteristics, they are treated as distinct datasets and modeled by independently trained neural networks. 

The sPHENIX calorimeter system comprises three sub-detectors~\cite{sphenix2017emcal}: an electromagnetic calorimeter (EMCal) and inner and outer hadronic calorimeters (IHCal, OHCal).  All three share a uniform, hermetic acceptance over pseudorapidity $-1.1 < \eta < 1.1$ and the full azimuth $0 < \phi < 2\pi$, and each is segmented into towers.  The EMCal towers are four times finer in each transverse direction, $\Delta\eta \times \Delta\phi = 0.025 \times 0.025$, than the $0.1 \times 0.1$ towers of the IHCal and OHCal.  To place the three calorimeters on a common grid, the energy in each block of $4 \times 4$ EMCal towers (sixteen towers) is aggregated and summed with the coincident single IHCal and OHCal towers, forming one integrated tower. This tiling yields $24$ integrated towers in $\eta$ and $64$ in $\phi$, and the transverse energy $\ET$ deposited in each is arranged into a single-channel image of shape $24 \times 64$ in the $(\eta,\phi)$ plane.  These per-event images constitute the ground-truth samples for training; approximately $6 \times 10^5$ events were used per centrality class for training.

\subsubsection{Network configuration and Training}
\label{sec:setup:sec2}

The authors of \cite{torbunov2024calo} used OpenAI's improved-DDPM (iDDPM) \cite{nichol2021improved} implementation for the noise-prediction network $\net(\bfx_t, t)$. Relative to the original DDPM formulation of \cite{ho2020ddpm}, iDDPM introduces several modifications (in addition to a mean, the variance of the reverse process is also made a learnable quantity which is optimized by a hybrid objective function that combines the KL divergence $D_{\rm K L}\left(q\left(\bfx_{t-1} \mid \bfx_t, \bfx_0\right) \| p_\theta\left(\bfx_{t-1} \mid \bfx_t\right)\right)$ for each diffusion time step with the simplified objective of \cite{ho2020ddpm} in Eq.~\eqref{eq:DDPM_objective}; further iDDPM introduces a cosine schedule for the cumulative noise level, from which the coefficients $\beta_t$ are determined). However the authors of \cite{torbunov2024calo} find that the fixed variance and linear variance schedule of \cite{ho2020ddpm} produce the best results. 

The underlying neural network for noise prediction $\net(x_t, t)$ is a convolutional U-Net~\cite{RonnebergerFB15} with three encoder-decoder stages of $32$, $64$, and $128$ channels, each stage containing two residual blocks~\cite{HeZRS15}. The scale-shift time modulation in the residual block was useful for only $40-50\%$ centrality dataset but not for $0-10\%$ centrality. In iDDPM's implementation there is also an option of up/down-sampling (per-resolution) attentions layers that are interleaved with the residual blocks. However, \cite{torbunov2024calo} found that these attention layers produced no appreciable improvement in the generation quality in either dataset and were therefore dropped (the self-attention mechanism in the bottleneck of the U-Net is still present). The diffusion schedule end point $\beta_T$ the $0$-$10\%$ centrality was taken to be $\beta_T = 0.0025$ while for the $40$-$50\%$ centrality, the model uses $\beta_T = 0.0125$\footnote{The values reported in the paper are not consistent with the values used in the Zenodo implementation which is what we quote above.}. The number of forward diffusion steps is fixed to be $T = 8000$. The U-Net parameters are optimized via Adam~\cite{kingma2014adam} with a constant learning rate ($10^{-4}$) and fixed batch size of $128$. Finally, the input data is normalized on a log scale after first clipping from below at $E_{\min} = 10^{-3}$~GeV (the minimum detector energy resolution), with samples returned to physical units by $\ET = e^{x}$.  The physical log-space range used by all clamps below is $[x_{\text{min}}, x_{\text{max}}] = [-7,4]$ ($\approx 0.9$ MeV to $55$ GeV per tower). All diffusion calculations - forward noising, denoising, and inpainting - take place in this clamped log representation.

\subsubsection{Sampling from the trained DDPM}
\label{sec:setup:sec3}

To sample from the trained DDPM the first step is to sample $\mathbf{x}_T \sim \mathcal{N}(\mathbf{0}, \mathbf{I})$\footnote{Strictly speaking, the terminal marginal distribution $q_T(\bfx_T)$ is obtained by averaging $q(\bfx_T\mid \bfx_0) = \mathcal{N}(\sqrt{\bar{\alpha}_T}\bfx_0, (1-\bar{\alpha}_T)\bfI)$ over the data distribution and is not generally Gaussian. However,  for $\bar{\alpha}_T \ll 1$, $q_T(\bfx_T) \simeq \mathcal{N}(\bfzero, \bfI)$ which motivates the standard initialization $ \bfx_T \sim \mathcal{N}(\bfzero, \bfI)$.}.

Then we take reverse steps from $t=T$ to $t=0$ by sampling $\mathbf{x}_{t-1}$ as follows
\begin{align}
\label{eq:ddpm-step}
    &\mathbf{x}_{t-1} = \ddpmstep(\mathbf{x}_t, \xhat)~,\\[5pt]
    &\ddpmstep(\mathbf{x}_t, \xhat) = \frac{1}{\sqrt{\alpha_t}} \left(\bfx_t-\frac{1-\alpha_t}{\sqrt{1-\bar{\alpha}_t}} \epsilon_\theta\left(\bfx_t, t\right)\right) \nonumber\\[5pt]
    &~+ \frac{\sqrt{1-\alpha_t}~\sqrt{1-\bar{\alpha}_{t-1}}}{\sqrt{1-\bar{\alpha}_t}} ~\bfz_t~, 
\label{eq:ddpm_def}
\end{align}
here $\bfz_t \sim \mathcal{N}(\mathbf{0}, \mathbf{I})$ is the fresh noise that is injected at every reverse step. The function definition $\ddpmstep(\mathbf{x}_t, \xhat)$ and its form in Eq.~\eqref{eq:ddpm_def} is derived in Appendix \ref{app:diffusion}. 

\begin{algorithm}[ht]
\caption{DDPM Sampling}
\label{alg:ddpm_sampling}
\begin{algorithmic}[1]

\State $\mathbf{x}_T \sim \mathcal{N}(\mathbf{0},\mathbf{I})$

\For{$t = T,\ldots,1$}

    \State $\epshat \gets \net(\bfx_t, t)$
    \State $\xhat \gets (\bfx_t - \sqrt{1-\bar{\alpha}_t}\,\epshat)/\sqrt{\bar{\alpha}_t}$
    \State
    $\displaystyle
    \mathbf{x}_{t-1}
    =
    \ddpmstep(\mathbf{x}_t, \xhat)
    $

\EndFor

\State \Return $\mathbf{x}_0$

\end{algorithmic}
\end{algorithm}

This is a full sampling procedure where the number of reverse denoising steps $S$ is the same as the number of forward noising steps $T=8000$. One can also do a sub-sampling where $S<T$ at the cost of fidelity but improved generation. In \cite{torbunov2024calo} it was found that sample generation quality improves monotonically as the number of reverse steps grows: $S=100, 200, 400, 600, 1000, 2000, 8000$ (for lower values of $S$ there is a significant divergence in the tail region of the energy distribution of the produced samples than from the data distribution). In the present work we set $S=T=8000$ for both event generation and inpainting study. This ensures that the sampler operates in the regime of highest fidelity and the inpainting algorithm performance we set out to measure is not confounded by sub-sampling artifacts\footnote{For the case of natural images, it was shown in ~\cite{song2021ddim} (commonly referred to as Denoising Diffusion Implicit Models (DDIMs)) that there is a principled way of sub-sampling from the trained DDPM network that works deterministically (i.e., for a given Gaussian random noise the reverse process always produces the same sample $\tilde{\mathbf{x}}_0$) and produces good samples. However in \cite{torbunov2024calo} it was shown that the DDIM sampling procedure leads to samples whose tail distribution is fatter than those that come from the DDPM sampler. The stochasticity of the DDPM reverse process appears to help correctly generate the low-statistics, high-energy regions.}. A sample of the energy plot for each centrality is shown in Figure \ref{fig:event_display}.

\section{Image Inpainting}
\label{sec:algorithms}

In this section we describe a Bayesian sampling approach for reconstructing missing or corrupted regions of a calorimeter data by viewing it as an image inpainting problem. After formulating the problem in the section below, we describe in detail some approaches for solving it. We provide the implementation details towards the end. For the Bayesian prior we use the pretrained unconditional diffusion model of \cite{torbunov2024calo} described in the previous section.

\subsection{Problem Formulation}
\label{sec:algorithms:ProblemFormulation}

Let $\mathbf{x}_0 \in \mathbb{R}^d$ denote the unobserved clean image, and let $M \in \{0,1\}^d$ be a binary observation mask, with $M_i=1$ on observed
(``live'') towers and $M_i=0$ on missing (``dead'') towers. If noise is ignored, then the image inpainting is the following linear inverse problem
\begin{equation}
\label{eq:def_inpainting}
    \mathbf{y} = M \odot \mathbf{x}_0 ,
\end{equation}
where $\bfy$ is the observed image, $M$ is the mask operator and $\odot$ is the element-wise (Hadamard) product. Equivalently, writing $H=\operatorname{diag}(M)$, the measurement model in Eq.~\eqref{eq:def_inpainting} can be written as $\bfy=H\bfx_0$. More generally, a linear inverse problem takes the form
\begin{equation}
    \bfy = H\bfx_0+\sigma_{\bfy}\boldsymbol{\xi},
    \qquad
    \boldsymbol{\xi}\sim\mathcal{N}(\mathbf{0},\mathbf{I}).
\end{equation}
When $H$ is rank deficient (as is the case of a mask operator), the problem is highly under-determined. The measurement alone does not uniquely determine $\bfx_0$. However, if the underlying distribution of $\bfx_0$ is known / approximately known, then a principled way of finding $\bfx_0$ is to draw it from the conditional distribution $p(\bfx_0\mid \bfy, M)$
\begin{equation}
\label{eq:conditional_distribution}
    p(\bfx_0\mid \bfy, M)
    \propto
    p(\bfy\mid\bfx_0, M)\,p(\bfx_0),
\end{equation}
where $p(\bfx_0)$ is the prior on $\bfx_0$. This point of view ensures two properties:
\begin{enumerate}
    \item The sampled image are \emph{data consistent} on the observed region,
    \item The procedure enforces \emph{realism/coherence} of the generated missing region under the prior $p(\mathbf{x}_0)$.
\end{enumerate}
In the case of noiseless measurements, every solution of \(H\bfx_0=\bfy\) can be decomposed into two orthogonal components: the range-space component of $H$ (these are the known / observed pixels) and the null space component of $H$ (the missing pixels). The measurement $\mathbf{y}$ constrains only the range-space component whereas the null space remains completely unconstrained. The general solution can be written as
\begin{equation}
    \bfx_0
    =
    H^\dagger\bfy
    +
    \bigl(\mathbf{I}-H^\dagger H\bigr)\bar{\bfx},
    \label{eq:app-nullspace}
\end{equation}
where $H^\dagger$ is the Moore-Penrose pseudoinverse that satisfies $HH^\dagger H = H$ and $\bar{\bfx}\in\mathbb{R}^d$ is arbitrary and must be determined by the prior. Therefore, in the noiseless case, the problem of finding the posterior distribution $p(\bfx_0 \mid \bfy) $ is simply the prior distribution restricted to the condition in Eq.~\eqref{eq:app-nullspace}.

For inpainting, $H=\operatorname{diag}(M)$ is an orthogonal
projector, so that
\begin{align}
    H=H^\top=H^2=H^\dagger .
\end{align}
The posterior of interest is therefore $p\!\left(\bfx_0^{\mathrm{dead}} \mid \bfx_0^{\mathrm{live}}=\bfy\right)$, the conditional law of the dead towers given the live ones under some known prior.

To summarize, the objective is not merely to produce one completion satisfying $M\odot\bfx_0=\bfy$, but to sample or estimate a plausible completion of $\mathbf{y}$ from the posterior distribution $p(\mathbf{x}_0 \mid \mathbf{y}, M)$.  A valid posterior must therefore satisfy two requirements mentioned above: exact data consistency on the live region and a faithful reconstruction of the dead region. 

\subsection{Diffusion-based inverse solvers}
\label{sec:algorithms:DiffusionSolvers}

A natural way to address the above problem is to use a pretrained unconditional diffusion model as a learned prior over clean images. In particular, DDPM can be conditioned on the measurement without retraining the denoising network. The inpainting algorithms that we consider in the ensuing sections differ primarily in how they impose the measurement constraint $M\odot \mathbf{x}_0=\mathbf{y}$ into the reverse diffusion process while keeping the unconditional denoising network fixed. RePaint \cite{lugmayr2022repaint} repeatedly replaces the observed region by a correctly noised version of the measurement and uses forward resampling to harmonize the boundary. DDNM \cite{wang2022ddnm} projects the predicted clean image onto the measurement-consistent affine subspace before each reverse step. DDRM \cite{kawar2022ddrm} performs the conditioning separately in the spectral range and null spaces of the degradation operator. The Gradient guided algorithms \cite{chung2022mcg, chung2023dps, song2023pigdm, grechka2024gradpaint, zirvi2025diffusion} instead modify the reverse dynamics through gradients of a measurement-consistency objective. 

Most of these algorithms have been designed for a larger class of problems that include noisy in-painting \cite{yeh2017semantic}, colorization \cite{larsson2016learning, zhang2016colorful}, image restoration \cite{kupyn2019deblurgan, suin2020spatially} and super-resolution \cite{haris2018deep, ledig2016photo}. However, because our interest in this work is to perform noise-free inpainting, we have specialized the discussion of these algorithms to the noise-free case. Table~\ref{tab:inpainters} summarizes the
algorithm-specific choices, costs, and consistency properties. 

\begin{table}[ht]
  \centering
  \setlength{\tabcolsep}{4pt}
  \begin{ruledtabular}
  \begin{tabular}{@{}llccc@{}}
    Algorithm & Params & Backprop & NFEs & s/image \\
    \colrule
    RePaint & $U = 10$ & no 
      & $U(S{-}1){+}1$ & $\approx 663$ \\
    DDNM & --- & no  & $S$ & $\approx 66$ \\
    DDRM & $\eta_a = 0.85$ & no 
      & $S$ & $\approx 66$ \\
    GGD & $g_s = 0.5$ & yes
      & $S$ ($+ S$ bwd) & $\approx 212$ \\
  \end{tabular}
  \end{ruledtabular}
    \caption{Summary of the implemented inpainting algorithms, specialized to noise-free case. ``Backprop'' indicates a backward pass through the $\net$-network per reverse diffusion step. NFEs are network function evaluations per posterior sample for $S$ reverse steps (`bwd' denotes additional backward passes). The rightmost column gives the measured wall-clock time per truth image (50 posterior samples, $S = T = 8000$, on the $8\times 8$ size mask).}
    \label{tab:inpainters}
\end{table}

\subsubsection{RePaint}
\label{sec:algorithms:repaint}

The RePaint algorithm \cite{lugmayr2022repaint} makes use of an unconditional diffusion model for the in-painting task by alternating an unconditional reverse
step on the dead region with an independent forward diffusion of the known content to the matching noise level and concatenating the 
\begin{align}
\label{eq:repaint-steps}
\begin{split}
  \mathbf{x}^{\mathrm{known}}_{t-1} &= \sqrt{\bar{\alpha}_{t-1}}\mathbf{y} + \sqrt{1-\bar{\alpha}_{t-1}} ~\epsilon~,\\
  \mathbf{x}^{\mathrm{dead}}_{t-1} &= \ddpmstep(\mathbf{x}_t, \xhat)~,\\
  \mathbf{x}'_{t-1} &= M \odot \mathbf{x}_{t-1}^{\mathrm{known}}
           + (1-M) \odot \mathbf{x}_{t-1}^{\mathrm{dead}} .
  \end{split}
\end{align}

Here $\ddpmstep(\mathbf{x}_t, \xhat)$ is one reverse DDPM step that is given by Eq.~\eqref{eq:ddpm-step}. The quantity $\mathbf{x}'_{t-1}$ again gets re-noised to noise level $t$ via the one-step forward schedule Eq.~\eqref{eq:fwd_process}. To harmonize the boundary between the in-painted dead region with the known surrounding region, the above steps in Eq.~\eqref{eq:repaint-steps} are repeated $U$ times ($U$ being a hyperparameter 
which Ref. \cite{lugmayr2022repaint} set to a default value of $U=10$). The noisy image at time step $t-1$ gets resampled $U$ times before taking one reverse DDPM transition from $\bfx_t \to \bfx_{t-1}$. At $t=1$ only a single backward pass is performed at which point $\bar{\alpha}_{0} = 1$ results in returning $\bfy$ exactly in the final step for the known region and a DDPM posterior sample filled in the masked region. The procedure is presented in Algorithm \ref{alg:repaint}. 

\begin{algorithm}[t]
\caption{RePaint}
\label{alg:repaint}
\begin{algorithmic}[1]
\State $\bfx_T \sim \mathcal{N}(0, \mathbf{I})$
\For{$t = T, \dots, 1$}
  \State $n_{\mathrm{inner}} \gets U$ \textbf{if} $t > 1$ \textbf{else} $1$
  \For{$u = 1, \dots, n_{\mathrm{inner}}$}
    \State $\epshat \gets \net(\bfx_t, t)$
    \State $\xhat \gets (\bfx_t - \sqrt{1-\bar{\alpha}_t}\,\epshat)/\sqrt{\bar{\alpha}_t}$
    \State $\bfx_{t-1}^{\mathrm{dead}} \gets \ddpmstep(\bfx_t, \xhat)$
    \State $\bfx^{\mathrm{known}}_{t-1} \gets \sqrt{\bar{\alpha}_{t-1}}\, \bfy
           + \sqrt{1-\bar{\alpha}_{t-1}}\;\epsilon$
    \State $\bfx'_{t-1} \gets M \odot \bfx_{t-1}^{\mathrm{known}}
           + (1-M) \odot \bfx_{t-1}^{\mathrm{dead}}$
    \If{$u < n_{\mathrm{inner}}$}
      \State $\bfx_t \gets \sqrt{\alpha_t}\, \bfx'_{t-1}
             + \sqrt{1-\alpha_t}\;\epsilon_t'$
    \Else
      \State $\bfx_{t-1} \gets \bfx'_{t-1}$
    \EndIf
  \EndFor
\EndFor
\State \Return $\bfx_0$
\end{algorithmic}
\end{algorithm}

\subsubsection{DDNM}
\label{sec:algorithms:ddnm}

The denoising diffusion null-space model (DDNM)~\cite{wang2022ddnm} is a sampling procedure that first corrects the known region of the denoised estimate before taking a reverse step. Concretely, for a given inverse problem $\bfy = H\bfx$, any solution $\bfx_0$ that is consistent with $\bfy = H \bfx_0$ can be decomposed into the range and null space of $H$. Written explicitly, $\bfx_0 = H^\dagger \bfy + (\mathbf{I} - H^\dagger H)\,\bar{\bfx}$ for any $\bar{\bfx}$. Here $H^\dagger$ is the Moore-Penrose pseudo inverse of $H$ that satisfies $HH^\dagger H = H$. For a noiseless mask, the projection of the denoised estimate $\xhat(\bfx_t)$ onto the range and null space of the masking operator is simply
\begin{equation}
  \xhat^{\,\mathrm{proj}}(\bfx_t) = M \odot \bfy + (1-M) \odot \xhat(\bfx_t)~.
  \label{eq:ddnm-proj}
\end{equation}
The projected denoised estimate $\xhat^{\,\mathrm{proj}}(\bfx_t)$ is then inserted into the usual reverse DDPM step $\ddpmstep(\bfx_t, \xhat^{\,\mathrm{proj}})$. By the collapse property of the final step, the output equals $\xhat^{\,\mathrm{proj}}$ at $t = 1$, so data consistency of the known live region is again exact. We note that in terms of code implementation DDNM is the minimal algorithm of all (one network evaluation and one projection per time step; there are no hyperparameters)

\begin{algorithm}[ht]
\caption{DDNM}
\label{alg:ddnm}
\begin{algorithmic}[1]
\State $\bfx_T \sim \mathcal{N}(\bfzero, \,\mathbf{I})$
\For{$t = T, \dots, 1$}
  \State $\epshat \gets \net(\bfx_t, t)$
  \State $\xhat \gets (\bfx_t - \sqrt{1-\bar{\alpha}_{t}}\,\epshat)/\sqrt{\bar{\alpha}_t}$
  \State $\xhat^{\,\mathrm{proj}} \gets M \odot \bfy + (1-M) \odot \xhat$
  \State $\bfx_{t-1} \gets \ddpmstep(\bfx_t, \xhat^{\,\mathrm{proj}})$
\EndFor
\State \Return $\bfx_0$
\end{algorithmic}
\end{algorithm}

\subsubsection{DDRM}
\label{sec:algorithms:ddrm}

Denoising diffusion restoration models (DDRM)~\cite{kawar2022ddrm} define a sampling procedure in the spectral space of the degradation operator $H$. For the case of noiseless inpainting, the spectral basis of the matrix $H$ coincides with the pixel basis. Setting $\sigma_\bfy = 0$ in the general update rule of the DDRM algorithm (Eqs.~(7) and (8) of Ref.~\cite{kawar2022ddrm}) and transforming it into the variance preserving variables used in this work, we get the following update rules
\begin{align}
  \bfx_{t-1}^{\rm{dead}} =& \sqrt{\bar{\alpha}_{t-1}}\,\xhat
        + \sqrt{(1-\eta_a^2)\,(1-\bar{\alpha}_{t-1})}\;\epsilon_\theta(\bfx_t, t)\nonumber\\
        &+ \eta_a \sqrt{1-\bar{\alpha}_{t-1}}\;\bfz_t ,
  \label{eq:ddrm-dead}\\[5pt]
  \bfx_{t-1}^{\rm{live}} =& \sqrt{\bar{\alpha}_{t-1}}\bigl((1-\eta_b)\,\xhat + \eta_b\, \bfy\bigr)
        + \sqrt{1-\bar{\alpha}_{t-1}}\;\bfz_t' .
  \label{eq:ddrm-obs}
\end{align}
In the above equations $\eta_a$ and $\eta_b$ are hyper-parameters that take values in the range $[0,1]$. Ref. \cite{kawar2022ddrm} used default values $\eta_a = 0.85$ and $\eta_b = 1$\footnote{A note on convention: $\eta_a$ here is referred to as $\eta$ in Ref. \cite{kawar2022ddrm}. In our work $\eta$ refers to pseudo-rapidity.}.  

The initialization of the reverse process in DDRM is conceptually different from the previous algorithms. At $t = T$, the live pixels start from $\mathcal{N}(\sqrt{\bar{\alpha}_T}\, \bfy,\, (1-\bar{\alpha}_{T})\,\mathbf{I})$ and the dead pixels from $\mathcal{N}(\bfzero, (1-\bar{\alpha}_{T})\,\mathbf{I})$. Then for all $t>1$, the pixels undergo updates according to the rule in Eqns. \eqref{eq:ddrm-dead} and \eqref{eq:ddrm-obs}. At the final reverse step ($t=1$) since $\bar{\alpha}_0 = 1$ the output live region exactly equals $\bfy$ (since $\eta_b = 1$) while the dead pixels are filled with the denoised estimate $\xhat$. We note that for $\eta_b < 1$ the last step would terminate at $(1-\eta_b)\,\xhat + \eta_b\, \bfy$ for the live region thereby leading to data inconsistency. We remark that Eq.~\eqref{eq:ddrm-dead} is DDIM-type~\cite{song2021ddim} update that becomes exactly the deterministic DDIM update rule for $\eta_a=0$ (see Appendix \ref{app:diffusion:posterior} for details). Among the in-painting algorithms that we study in this work, DDRM is the only algorithm that implements a DDIM-type transport. Note that, as opposed to the general update rule for DDIM (which reduces to the usual DDPM update rule for a specific choice of the variance), for the DDRM construction there is no choice of ($t$-independent) $\eta_a$ for which the sampling procedure reduces to that of DDPM update rule.

\begin{algorithm}[t]
\caption{DDRM (noise-free)}
\label{alg:ddrm}
\begin{algorithmic}[1]
\State $\bfx_T \gets M \odot \bigl(\sqrt{\bar{\alpha}_T}\, \bfy + \sqrt{1-\bar{\alpha}_T}\,\bfz_T\bigr)
        + (1-M) \odot \sqrt{1-\bar{\alpha}_T}\,\bfz_T$
\For{$t = T, \dots, 1$}
    \State $\epshat \gets \net(\bfx_t, t)$
  \State $\xhat \gets (\bfx_t - \sqrt{1-\bar{\alpha}_{t}}\,\epshat)/\sqrt{\bar{\alpha}_t}$
  \If{$t > 1$}
    \State $\begin{aligned}[t]
  \bfx_{t-1}^{\mathrm{dead}} \gets{}\; & \sqrt{\bar{\alpha}_{t-1}}\,\xhat
        + \sqrt{(1-\eta^2_a)\,(1-\bar{\alpha}_{t-1})}\;\epshat \\
      & {}+ \eta_a\sqrt{1-\bar{\alpha}_{t-1}}\;\bfz_t
\end{aligned}$
\vspace{3pt}
    \State $\bfx_{t-1}^{\mathrm{live}} \gets \sqrt{\bar{\alpha}_{t-1}}\bigl[(1-\eta_b)\,\xhat
           + \eta_b\, \bfy\bigr] + \sqrt{1-\bar{\alpha}_{t-1}}\;\bfz_t'$
  \Else
    \State $\bfx^{\mathrm{dead}}_{t-1} \gets \xhat$
    \State $\bfx^{\mathrm{live}}_{t-1} \gets \bfy$
           %\Comment{final projection; no-op at $\eta_b = 1$}
  \EndIf
  \State $\bfx_{t-1} \gets M \odot \bfx_{t-1}^{\mathrm{live}} + (1-M) \odot \bfx_{t-1}^{\mathrm{dead}}$
\EndFor
\State \Return $\bfx_0$
\end{algorithmic}
\end{algorithm}

\subsubsection{Gradient-guided in-painting algorithms}
\label{sec:algorithms:guided}

The algorithms we discussed so far follow the reverse diffusion process and impose consistency with the measurement, by projecting the known pixels at every step of the denoising process. Gradient-guided in-painting algorithms \cite{chung2022mcg, chung2023dps, song2023pigdm, grechka2024gradpaint} supplement this sampling procedure with an additional gradient guided step \cite{dhariwal2021diffusion, ho2022classifier, ho2022video} that we describe in-brief.

At each reverse step, we first use the unconditional denoiser to form a clean estimate $\xhat\left(\bfx_t, t\right)$. We then calculate the measurement loss $\mathcal{L}_t$ defined as the $L^2$ norm of the difference between the denoised estimate $\xhat(\bfx_t, t)$ and the measurement $\bfy$ for the known pixels
\begin{align}
\label{eq:guided-loss}
  \mathcal{L}_t = 
  \bigl\lVert M \odot (\xhat\left(\bfx_t, t\right) - \bfy) \bigr\rVert_2^2~.
\end{align}
The gradient of this measurement loss $\mathbf{g}_t$ is the following Jacobian-vector product
\begin{align}
    \mathbf{g}_t & = \nabla_{\bfx_t} \mathcal{L}_t \nonumber \\
    & = 2[\nabla_{\bfx_t} \xhat\left(\bfx_t, t\right)]^\top [M \odot (\xhat\left(\bfx_t, t\right) - \bfy) ]~.
\end{align}
This quantity is computed by backpropagating the measurement loss through the denoiser. The computed gradient provides a guidance direction in the current noisy state where the usual DDPM reverse step acquires a correction term proportional to the gradient of the measurement loss 
\begin{align}
\label{eq:guided-step}
\bfx_{t-1}^{\mathrm{guided}} = \ddpmstep(\bfx_t, \xhat) - \lambda_t\, \mathbf{g}_t    
\end{align}
The rest of the algorithm is a projection step (common to both RePaint and DDRM) where the known region is noised to the appropriate noise level, while the missing region is taken from the gradient guided reverse diffusion step. The complete procedure is described in Algorithm \ref{alg:guided}. 
\begin{algorithm}[t]
\caption{Gradient Guided Diffusion}
\label{alg:guided}
\begin{algorithmic}[1]
\State $\bfx_T \sim \mathcal{N}(0,\mathbf{I})$
\For{$t = T, \dots, 1$}
  \State \textbf{with} gradients enabled:
  \State $\epshat \gets \net(\bfx_t, t)$
  \State $\xhat(\bfx_t) \gets (\bfx_t - \sqrt{1-\bar{\alpha}_{t}}\,\epshat)/\sqrt{\bar{\alpha}_{t}}$
  \State $\mathbf{g}_t \gets \nabla_{\bfx_t} \lVert M \odot (\xhat(\bfx_t) - \bfy)\rVert_2^2$
  \State $\bfx_{t-1}^{\mathrm{guided}} \gets \ddpmstep(\bfx_t, \xhat) - \lambda_t\, \mathbf{g}_t$
  \State $\bfx_{t-1}^{\mathrm{known}} \gets \sqrt{\bar{\alpha}_{t-1}}\, \bfy + \sqrt{1-\bar{\alpha}_{t-1}}\;\epsilon$
  \State $\bfx_{t-1} \gets M \odot \bfx_{t-1}^{\mathrm{known}} + (1-M) \odot \bfx_{t-1}^{\mathrm{guided}}$
\EndFor
\State \Return $\bfx_0$
\end{algorithmic}
\end{algorithm}
The quantity $\lambda_t$ in Eq.~\eqref{eq:guided-step} is a tunable guidance weight which we set to 
\begin{align}
\lambda_t = g_s(1-\bar{\alpha}_t)~,~~~~g_s = \frac{1}{2}~.
\end{align}
such that it anneals with the marginal variance ($\lambda_t$ is large in the region $t\to T$ and vanishes as $t\to 0$). An important implementation detail is that gradient guidance through the diffusion model is necessarily numerically unstable on long schedules. The de-noised estimate
\begin{align}
\label{eq:denoised-estimate}
\xhat(\bfx_t)=
\frac{\bfx_t-\sqrt{1-\bar\alpha_t}\,\epsilon_\theta(\bfx_t,t)}
{\sqrt{\bar\alpha_t}}
\end{align}
amplifies denoising errors by \(1/\sqrt{\bar\alpha_t}\) at high noise levels. Hence the finiteness of its derivative
\begin{align}
\label{eq:denoised-estimate-gradient}
\nabla_{\bfx_t}\xhat=
\frac{
\bfI-\sqrt{1-\bar\alpha_t}\nabla_{\bfx_t} \epsilon_\theta(\bfx_t, t)
}
{\sqrt{\bar\alpha_t}},
\end{align}
relies on the numerator in Eq.~\eqref{eq:denoised-estimate-gradient} being at least $O(\sqrt{\bar{\alpha}_t})$. In Appendix \ref{app:inverse:optimal-denoiser} we show that the gradient of the optimal denoiser\footnote{For an ideal denoising network we have $\net^*(\bfx_t, t) = \mathbb{E}[\epsilon \mid \bfx_t]$ or equivalently $\bfx_0^*\left(\bfx_t, t\right)=\mathbb{E}\left[\bfx_0 \mid \bfx_t\right]$ where the expectation is taken over the data distribution \cite{vincent2011connection, efron2011tweedie}.} is $O(\sqrt{\bar{\alpha}_t})$ and therefore $\nabla_{\bfx_t} \epsilon_\theta(\bfx_t, t) = \bfI + O(\bar{\alpha}_t)$. In practice however, the $\epsilon_\theta$-network will generally produce a schedule agnostic $O(1)$ error term in its gradient that can come for instance from model misspecification error and numerical precision. Consequently, the unregularized loop from computing $\xhat$ to the gradient of the measurement loss and back to $\bfx_t$ results in a floating-point overflow in just a few reverse steps. Hence some sort of regularization is needed. In our implementation we regularize the loop by clipping the denoised estimate $\xhat^c = \operatorname{clip}(\xhat, x_{\min}, x_{\max})$ where $x_{\min}=-7, x_{\max}=4$ corresponds to the physical region in log space (c.f. discussion in Section \ref{sec:setup:sec2}).

\subsection{Inpainting methodology}
\label{sec:algorithms:calo-inpainting}

This Section presents some details of our implementation. As discussed in Section \ref{sec:calo-ddpm-review} calorimeter images take the form of $(24,64)$ pixel values. In this image a dead detector region is essentially a binary mask $M \in \{0,1\}^{24\times 64}$ with $M_i = 1$ on live towers and $M_i = 0$ on dead towers. The live towers are measured with the detector's ordinary resolution (which is part of what the generative model was trained to describe) while the dead towers have zero pixel values which are filled in using the various diffusion based posterior sampling methods described in the previous section. 

For every algorithm sampling is performed by batching $J=50$ independent samples per truth image with $K=20$ truth images giving a total GPU batch
size $KJ = 1000$. In total we analyzed $1000$ truth images. Each algorithm inputs one measurement pair $(\bfy, M)$ and returns $J$ samples of the full in-painted image. As for the geometry of the dead region, the methods of the previous section can handle any shape in principle\footnote{The authors of DDNM~\cite{wang2022ddnm} report that its perceptual realism can deteriorate for difficult inverse problems, including inpainting with large missing regions. To improve global coherence of the inpainting, they introduce DDNM+ algorithm with a RePaint-inspired ``time-travel'' procedure that alternates forward noising with repeated reverse DDNM steps.}. In our study, we worked with a square $b\times b$ mask at a fixed position $(\eta_0, \phi_0) = (8, 28)$, with $b \in \{6, 8, 10, 12\}$. For comparing various algorithms we picked $b=8$. All implementations share the same variance schedule obtained from the pre-trained diffusion model's configuration. We do not perform subsampling and $S=T=8000$ for all experiments.

\section{Results and discussion}
\label{sec:results}

\begin{figure*}[ht]
\centering
\includegraphics[width=0.9\textwidth]{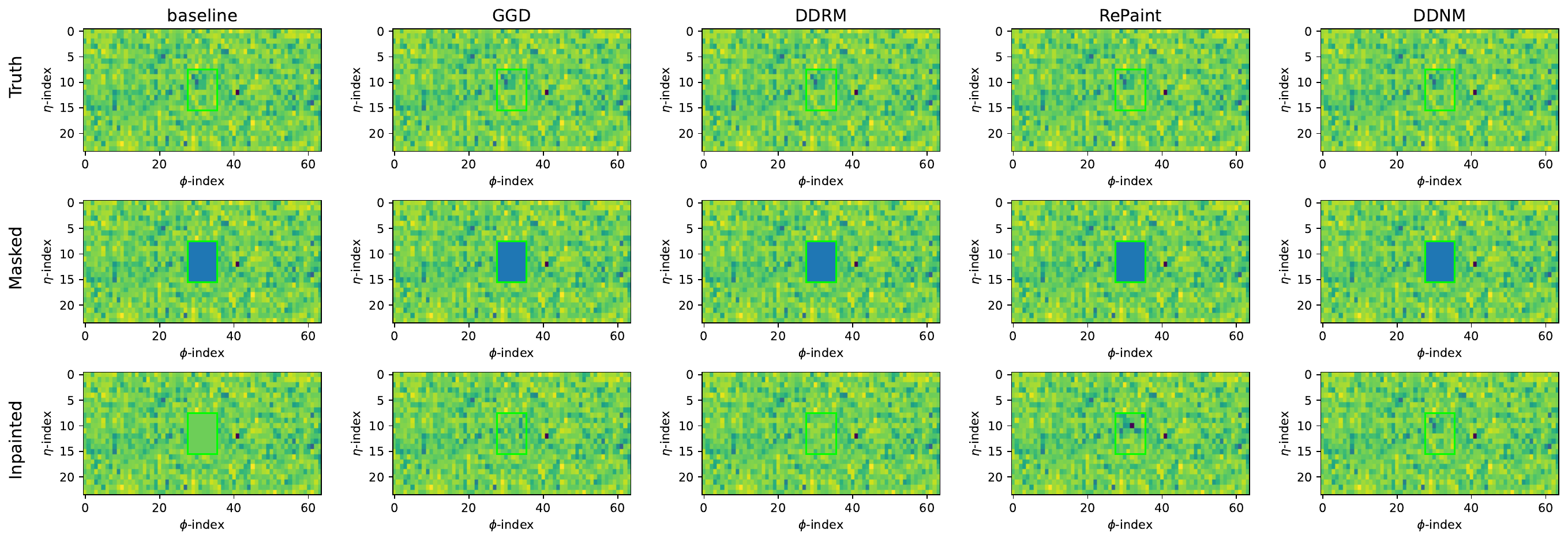}
\caption{Representative examples of reconstructed calorimeter images for different inpainting algorithms. The comparison includes the ground-truth event (top row), the masked input (middle row), and a representative posterior sample (bottom row). The box shows the region which is being inpainted.}
  \label{fig:examples}
\end{figure*}

In our study, we sample an ensemble of $N=1000$ images $\bfx^{\rm true}_i$, ($i=1,\cdots N$) from the pre-trained unconditional calorimeter DDPM of \cite{torbunov2024calo} which are viewed as the ground truth. For each ground truth image $\bfx^{\rm true}_i$, ($i=1,\cdots N$) we construct a masked measurement $\bfy_i = M \odot\bfx^{\rm true}_i$ and use the same DDPM as the prior for inpainting conditioned on $(\bfy_i, M)$. Under this setup, an exact posterior sampler would make the ground-truth image and the reconstructed posterior samples conditionally exchangeable given $(\bfy_i, M)$. Any systematic deviation of the diagnostics from their expected behavior, beyond finite-sample fluctuations, can therefore be attributed to algorithmic or numerical approximations.

The inpainting algorithms are evaluated using statistical metrics for assessing both reconstruction accuracy and probabilistic calibration. Rather than relying solely on pixel-wise differences between generated samples and the ground truth, the evaluation includes measurements of global energy reconstruction, spatial bias, and other distributional properties. All tests are performed on an event-by-event basis using the complete posterior ensemble generated by each algorithm.

As a baseline for dead pixel fills, we use a constant mean fill with mean taken over the observed pixels. The performance of four algorithms discussed in Section~\ref{sec:algorithms} are compared against this baseline on a square mask of size of $8\times8$. In addition, two different event multiplicity classes are compared to understand the effect of event activity using the DDNM algorithm. Finally, a box-size sweep from $6\times6$ to $12\times12$ is performed to investigate the dependence on the size of the dead region. 

Representative examples of reconstructed calorimeter images for different inpainting algorithms are presented in Figure \ref{fig:examples}. The comparison includes the ground-truth event, the masked input, and a representative posterior sample from the different inpainting algorithms. The reconstructed events faithfully reproduce the principal shower topology.

{\it Notation}: For this Section the subscript $i$ on $\bfx_i^{\text{true}}$ will denote the images in the ensemble, $x_{i,p}^{\text{true}}$ will be used to denote the pixel $p$ of $\bfx_i^{\text{true}}$. The notation $\bfx_i^{(j)}$ for $j=1,\cdots J$ will be used for the $J=50$ posterior fills with $x_{i,p}^{(j)}$ denoting the $p$'th pixel of $\bfx_i^{(j)}$. The region $\mathcal C$ will denote the set of all pixels whereas the region $\mathcal{D}$ will denote the set of masked pixels.

%%%%%%%%%%%%%%%%%%%%%%%%%%%%%%%%%%%%%%%%%%%%%%%%%%%%%%%%%%%%%%
%%%%%%%%%%%%%%%%%%%%%%%%%%%%%%%%%%%%%%%%%%%%%%%%%%%%%%%%%%%%%%

\subsection{Comparison of inpainting algorithms}
\label{sec:results:inpainting-comparison}
The performance of four different inpainting algorithms are compared against the mean fill baseline using various statistical measures for a box-size of $8\times8$ and events with high multiplicities.

%%%%%%%%%%%%%%%%%%%%%%%%%%%%%%%%%%%%%%%%%%%%%%%%%%%%%%%%%%%%%%
%%%%%%%%%%%%%%%%%%%%%%%%%%%%%%%%%%%%%%%%%%%%%%%%%%%%%%%%%%%%%%

\subsubsection{Global energy reconstruction}
\label{sec:results:inpainting-comparison:global}
An important consistency check is whether the reconstructed images preserve the total event energy. For event $i$, the total true energy and the reconstructed energy of
the $j$th posterior sample are
\begin{equation}
E_i^{\mathrm{true}}
=
\sum_{p\in\mathcal C}x_{i,p}^{\mathrm{true}},
\qquad
E_i^{(j)}
=
\sum_{p\in\mathcal C}x_{i,p}^{(j)} .
\end{equation}
The energy response ratio is defined by
\begin{equation}
R_i^{(j)}=\frac{E_i^{(j)}}{E_i^{\mathrm{true}}}.
\end{equation}
For each event, the posterior mean response and its standard deviation are
estimated from the posterior samples 
\begin{equation}
\bar R_i
=
\frac{1}{J}\sum_{j=1}^{J}R_i^{(j)}~,~~
s_{R,i}^2
=
\frac{1}{J-1}
\sum_{j=1}^{J}
\left(R_i^{(j)}-\bar R_i\right)^2~.
\end{equation}

Values of $\bar R_i$ close to unity indicate that the reconstruction preserves the global event energy, while $s_{R,i}$ quantifies the posterior uncertainty in the energy response. Deviations from unity primarily probe systematic biases introduced by the inpainting procedure rather than the missing region itself.
\begin{figure}[h]
\centering
  \includegraphics[width=0.95\columnwidth]{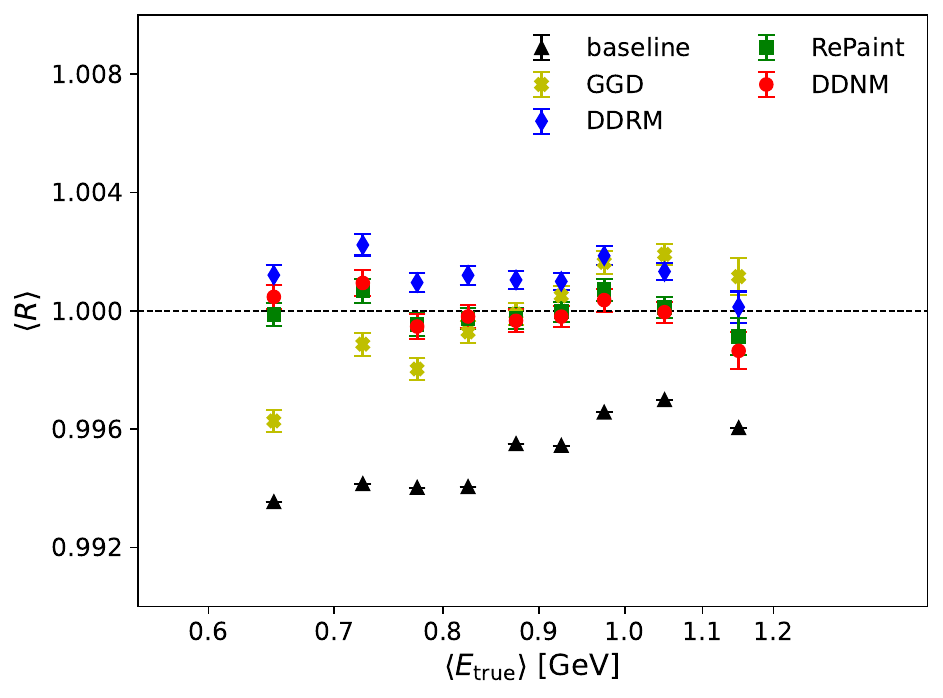}
  \caption{The ratio of reconstructed to full energy as a function of the average true energy for different inpainting methods. }
  \label{fig:full_ratio}
\end{figure}
Figure~\ref{fig:full_ratio} shows that the various methods preserve the total event energy to a high degree of accuracy. The small deviations from unity across the full range of event energies, particularly for the RePaint and DDNM methods, indicate that these reconstruction procedures do not introduce significant global normalization biases. The baseline, on the other hand, systematically underestimates the energy response.

%%%%%%%%%%%%%%%%%%%%%%%%%%%%%%%%%%%%%%%%%%%%%%%%%%%%%%%%%%%%%%
%%%%%%%%%%%%%%%%%%%%%%%%%%%%%%%%%%%%%%%%%%%%%%%%%%%%%%%%%%%%%%

\subsubsection{Posterior mean shrinkage}
\label{sec:results:inpainting-comparison:shrinkage}

\begin{table*}[ht]
  \centering
  \setlength{\tabcolsep}{4pt}
  \begin{ruledtabular}
  \begin{tabular}{@{}lccccc@{}}
   & \multicolumn{2}{c}{Mean}
  & & \multicolumn{2}{c}{Intercept}\\
\cline{2-3} \cline{5-6}
    Model & True & Masked & Slope & Expected & Obtained \\
    \colrule
    Baseline & 0.712$\pm$0.002 & 0.648$\pm$0.002 & 0.75$\pm$0.02 & 0.18$\pm$0.01 & 0.12$\pm$0.01\\    
    GGD & 0.712$\pm$0.002 & 0.708$\pm$0.009 & 0.83$\pm$0.02 & 0.12$\pm$0.01 & 0.12$\pm$0.01 \\
    DDRM & 0.712$\pm$0.002 & 0.739$\pm$0.008 & 0.66$\pm$0.02 & 0.24$\pm$0.01 & 0.27$\pm$0.01 \\
    RePaint & 0.712$\pm$0.002 & 0.713$\pm$0.009 & 0.65$\pm$0.02 & 0.25$\pm$0.01 & 0.25$\pm$0.01 \\
    DDNM & 0.712$\pm$0.002 & 0.711$\pm$0.009 & 0.61$\pm$0.02 & 0.28$\pm$0.01 & 0.28$\pm$0.01 \\
   \end{tabular}
  \end{ruledtabular}
    \caption{Parameters of the forward shrinkage study for different inpainting models. The true and masked mean respectively estimate $\mu$ and $\mathbb{E}[m]$. The expected intercept is given by $(1- \mathrm{slope})*\mathrm{true \, mean}$.}
    \label{tab:method_shrinkage_forward}
\end{table*}

\begin{figure*}[!htb]
\centering
  \includegraphics[width=\textwidth]{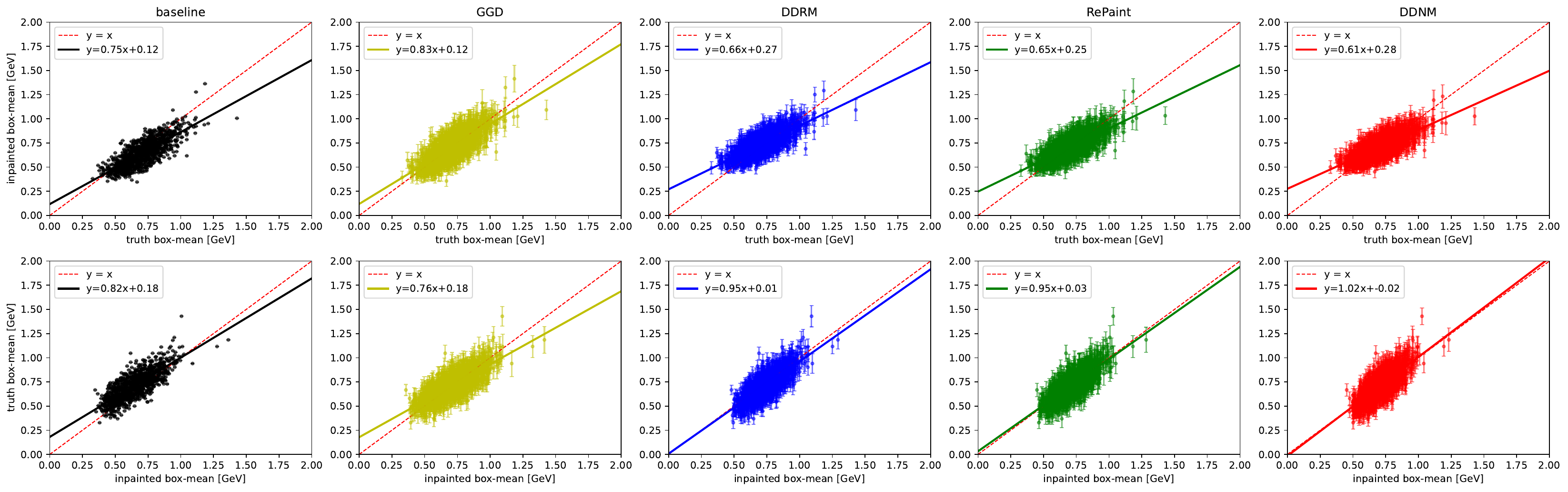}
  \caption{Forward shrinkage of mean reconstructed energy towards true missing energy for different methods is shown in the top panel. The reverse shrinkage, true missing energy towards reconstructed energy is shown on the bottom panel. The slope and intercept of the fit is also included.}
  \label{fig:shrinkage}
\end{figure*}

Mean shrinkage is a well-known feature of Bayesian models (see for instance Chapter 2 of \cite{gelman2013bda}). It measures how faithfully each algorithm captures properties of the posterior distribution in Eq.~\eqref{eq:conditional_distribution}. Consider an image $\bfx_i^{\text{true}}$ ($i=1, \cdots, N$) out of the set of $N=1000$ images. Denote the non-masked region of $\bfx_i^{\text{true}}$ by $\bfx_{i, {\rm known}}$ and define the following quantities 
\begin{align}
\label{eq:true_T}
    T_i &= \frac{1}{|\mathcal{D}|}\sum_{p\in\mathcal{D}} x_{i,p}^{\text{true}}~,~~
    \mu = \mathbb{E}[T]~,\\[5pt]
\label{eq:m_i}
    m_i &= \frac{1}{|\mathcal{D}|}\sum_{p\in\mathcal{D}}\mathbb{E} [x_{i,p} | \bfx_{i, {\rm known}}]~.
\end{align}
Here $T_i$ is the patch mean of the image $\bfx_i^{\text{true}}$ in the masked region $\mathcal{D}$. $\mu$ is the global patch mean across all true images and $m$ is the posterior mean of the patch mean conditioned on the non-masked region. By the iterative property of conditional expectations the quantities $\mu$ and $m_i$ are related by $\mathbb{E}[m]=\mu$. The expectation in the definition of $\mu$ can be estimated using the $N$ true samples: $\mu\approx \frac{1}{N}\sum_{i=1}^N T_i$. The conditional expectation in $m_i$ can be estimated by averaging the filled in patch mean over the $J=50$ generated samples:
\begin{align}
    \widehat m_i = \frac{1}{|\mathcal{D}|}\sum_{p\in\mathcal{D}} \frac{1}{J}\sum_{j=1}^J x_{i,p}^{(j)}~.
\end{align}
In general the true patch mean $T$ will not be equal to the conditional mean $m$ of the generated samples. They differ by fluctuations 
\begin{align}
\label{eq:T-m-fluctuations}
    T = m + \varepsilon~,\qquad \mathbb{E}[\varepsilon | \bfx_{{\rm known}}] = 0~.
\end{align}
Note that $m$ and $\varepsilon$ are un-correlated: ${\rm Cov}(m,\varepsilon) = \mathbb{E}[m\varepsilon]=\mathbb{E}[\mathbb{E}[m\varepsilon | \bfx_{{\rm known}}]]=\mathbb{E}[m\mathbb{E}[\varepsilon | \bfx_{{\rm known}}]]=0$. Therefore we have ${\rm Var}(T) = {\rm Var}(m) + {\rm Var}(\varepsilon)$ and ${\rm Cov}(m, T) ={\rm Var}(m) $. Consequently, the least square slope $s$ of $m$ against $T$ is
\begin{align}
    s = \frac{{\rm Cov}(m, T)}{{\rm Var}(T)} = \frac{{\rm Var}(m)}{{\rm Var}(m) + {\rm Var}(\varepsilon)} < 1~,
\end{align}
where the inequality follows since ${\rm Var}(\varepsilon)>0$ in general. Furthermore, the least square intercept $b = \mathbb{E}[m] - s\mathbb{E}[T] = (1-s)\mu$ is also constrained by the slope and therefore provides an independent consistency check. A further consistency check comes from regressing $T_i$ against $m_i$ in which case the least square slope is
\begin{align}
    s = \frac{{\rm Cov}(m, T)}{{\rm Var}(m)} = 1
\end{align}
and the intercept is $b = \mathbb{E}[T] - s\mathbb{E}[m] = 0$. 
\begin{table}[ht]
  \centering
  \setlength{\tabcolsep}{4pt}
  \begin{ruledtabular}
  \begin{tabular}{@{}lcc@{}}
    Model & Slope & Intercept \\
    \colrule
    Baseline & 0.82$\pm$0.02 & 0.18$\pm$0.01\\    
    GGD & 0.76$\pm$0.02 & 0.18$\pm$0.01 \\
    DDRM & 0.95$\pm$0.02 & 0.01$\pm$0.02 \\
    RePaint & 0.95$\pm$0.02 & 0.03$\pm$0.02 \\
    DDNM & 1.02$\pm$0.03 & -0.02$\pm$0.02 \\
   \end{tabular}
  \end{ruledtabular}
    \caption{Measured parameters of the reverse shrinkage study for different inpainting models. The expected slope and intercept are 1 and 0, respectively.}
    \label{tab:method_shrinkage_reverse}
\end{table}
These expectations are tested in Figure \ref{fig:shrinkage} and in Tables \ref{tab:method_shrinkage_forward} and \ref{tab:method_shrinkage_reverse}. The first and second columns in the Table ~\ref{tab:method_shrinkage_forward} are $\mu$ and $\mathbb{E}[m]$ which should be equal for an ideal posterior sampler. The third column is the slope $s$ of the fitted regression line which gives a prediction for what the regression intercept in the fourth column should be. The actual intercept obtained from the regression fit is written in the last column.  In the top row of Figure \ref{fig:shrinkage} we regress the box-mean of the posterior samples averaged over all the generated samples, $m_i$, against the box-mean of truth, $T_i$. The reverse regression is performed in the bottom row and the measured values are given in Table~\ref{tab:method_shrinkage_reverse}. While all methods are observed to recover the overall correlation between the reconstructed and true energies, they differ in the extent to which they reproduce the full dynamic range. The forward slopes range from $0.61$ to $0.83$, reflecting the expected shrinkage of the posterior mean toward the prior expectation. The reverse regression provides a more direct test of calibration: DDRM and RePaint both give slopes of $0.95\pm0.02$, while DDNM gives $1.02\pm0.03$, with intercepts consistent with their expected values within the uncertainties. In contrast, the baseline and GGD methods show larger deviations from unit slope. These results indicate that DDRM, RePaint, and DDNM reproduce the event-by-event variation of the missing energy more faithfully, with DDNM giving the closest agreement with the expected unit-slope calibration.

%%%%%%%%%%%%%%%%%%%%%%%%%%%%%%%%%%%%%%%%%%%%%%%%%%%%%%%%%%%%%%
%%%%%%%%%%%%%%%%%%%%%%%%%%%%%%%%%%%%%%%%%%%%%%%%%%%%%%%%%%%%%%

\subsubsection{Spatial bias}
\label{sec:results:inpainting-comparison:bias}

To investigate localized reconstruction biases, we compute the average residual over the complete sample space $\Delta(\eta,\phi) \equiv \Delta_p$ as follows. For a pixel $p\in\mathcal D$ in the masked region, we define the
pixel-wise conditional mean
\begin{equation}
m_{i,p}
\equiv
\mathbb{E}\!\left[
x_{i,p}\mid\mathbf{x}_{i,\mathrm{known}}
\right],
\end{equation}
which is estimated from the $J$ posterior samples as
\begin{equation}
\label{eq:m_hat}
\widehat m_{i,p}
=
\frac{1}{J}
\sum_{j=1}^{J}
x_{i,p}^{(j)}.
\end{equation}
The spatial reconstruction bias is then
\begin{equation}
\Delta_p
=
\frac{1}{N}
\sum_{i=1}^{N}
\left(
\widehat m_{i,p}
-
x_{i,p}^{\mathrm{true}}
\right),
\qquad p\in\mathcal D .
\end{equation}

\begin{figure*}[htb]
\centering
  \includegraphics[width=\textwidth]{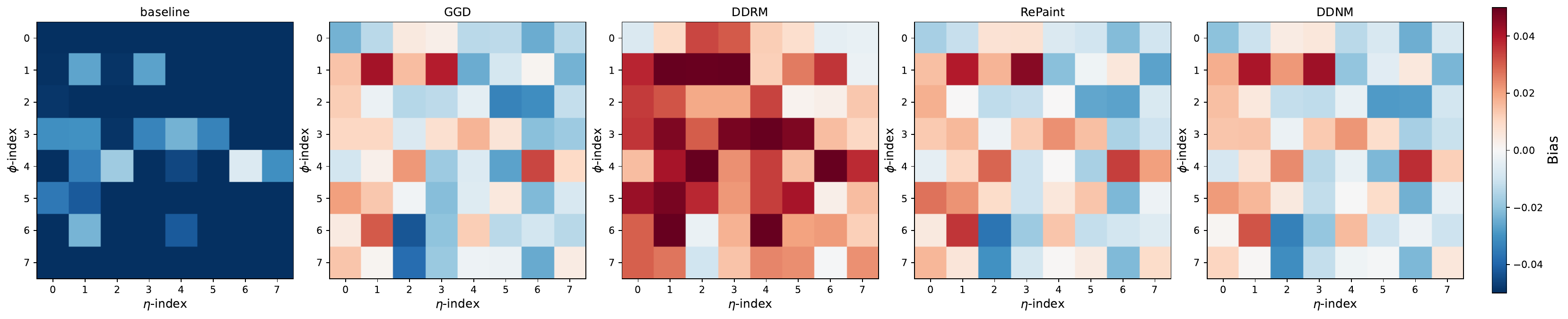}
  \caption{Average residual as a function of $\eta-$ and $\phi-$index for different algorithms.}
  \label{fig:biasmaps}
\end{figure*}

The resulting spatial bias maps are presented in Figure~\ref{fig:biasmaps}. An unbiased reconstruction should produce residuals centered around zero without visible detector-dependent structures. Localized positive or negative residuals would indicate systematic artifacts introduced by the inpainting procedure. The baseline systematically underestimates the reconstructed energy, while DDRM shows a systematic overestimate, consistent with the trends observed in the total energy response. In contrast, the residuals for GGD, RePaint, and DDNM remain small across the phase space, with no significant position-dependent structures. These results indicate that these methods do not introduce appreciable localized reconstruction biases.

%%%%%%%%%%%%%%%%%%%%%%%%%%%%%%%%%%%%%%%%%%%%%%%%%%%%%%%%%%%%%%
%%%%%%%%%%%%%%%%%%%%%%%%%%%%%%%%%%%%%%%%%%%%%%%%%%%%%%%%%%%%%%

\subsubsection{Posterior consistency and probabilistic calibration}
\label{sec:results:inpainting-comparison:calibration}

Because diffusion-based reconstruction produces an ensemble of possible solutions rather than a single deterministic image, it is important to verify that the predicted uncertainty reflects the true statistical fluctuations. The probabilistic performance of the reconstruction is evaluated using the standardized residual ($z$-score) and the Probability Integral Transform (PIT). These complementary diagnostics assess whether the distribution predicted by the diffusion model is statistically consistent with the true calorimeter energy.

The $z$-score measures how far the true value lies from the predicted mean in units of the predicted uncertainty. We calculate it as follows. For the $j$th posterior sample of image $i$, let
\begin{equation}
T_i^{(j)}
=
\frac{1}{|\mathcal D|}
\sum_{p\in\mathcal D}
x_{i,p}^{(j)}
\end{equation}
denote the patch mean energy in the masked region $\mathcal{D}$. The conditional mean
$m_i=\mathbb{E}[T_i\mid\mathbf{x}_{i,\mathrm{known}}]$ and its posterior
standard deviation are estimated from the $J$ samples as
\begin{equation}
\widehat m_i
=
\frac{1}{J}
\sum_{j=1}^{J}T_i^{(j)},
\qquad
\widehat\sigma_i^2
=
\frac{1}{J-1}
\sum_{j=1}^{J}
\left(T_i^{(j)}-\widehat m_i\right)^2
~.
\end{equation}
The corresponding event-level standardized residual is
\begin{equation}
\label{eq:event_level_zscore}
z_i=\frac{T_i-\widehat m_i}{\widehat\sigma_i}~,
\end{equation}
where $T_i$ was defined in Eq.~\eqref{eq:true_T}. The expected $z$-score and its standard deviation are then defined by
\begin{align}
    \langle z_\mathrm{score}\rangle = \frac{1}{N}\sum_{i=1}^N z_i~,~\sigma_{z_\mathrm{score}}^2 = \frac{1}{N-1}\sum_{i=1}^N\left(z_i - \langle z_\mathrm{score}\rangle\right)^2~.
\end{align}

For a well-calibrated reconstruction, the $z$-scores should follow a standard normal distribution with mean zero and unit variance, indicating that the reconstruction is unbiased and that the estimated uncertainties have the correct magnitude. The measured $z$-score for various algorithms is compiled in Table~\ref{tab:modelzsharpness}. A nonzero mean indicates a systematic bias in the posterior mean, while a width different from unity indicates under- or overestimation of the predictive uncertainty.

\begin{table}[ht]
  \centering
  \setlength{\tabcolsep}{4pt}
  \begin{ruledtabular}
  \begin{tabular}{@{}lccc@{}}
    Model & $\langle z_\mathrm{score}\rangle$ & $\sigma_{z_\mathrm{score}}$ & sharpness \\
    \colrule
        
       GGD & 0.11$\pm$0.04 & 1.21$\pm$0.03 & 0.496$\pm$0.003 \\
        DDRM & -0.36$\pm$0.04 & 1.24$\pm$0.03 & 0.433$\pm$0.003 \\
        RePaint & 0.01$\pm$0.03 & 1.09$\pm$0.02 & 0.495$\pm$0.003 \\
        DDNM & 0.01$\pm$0.03 & 1.03$\pm$0.02 & 0.497$\pm$0.003 \\
         
   \end{tabular}
  \end{ruledtabular}
    \caption{Model dependence of $z-$score and sharpness}
    \label{tab:modelzsharpness}
\end{table}

\begin{figure*}[ht]
\centering
  \includegraphics[width=\textwidth]{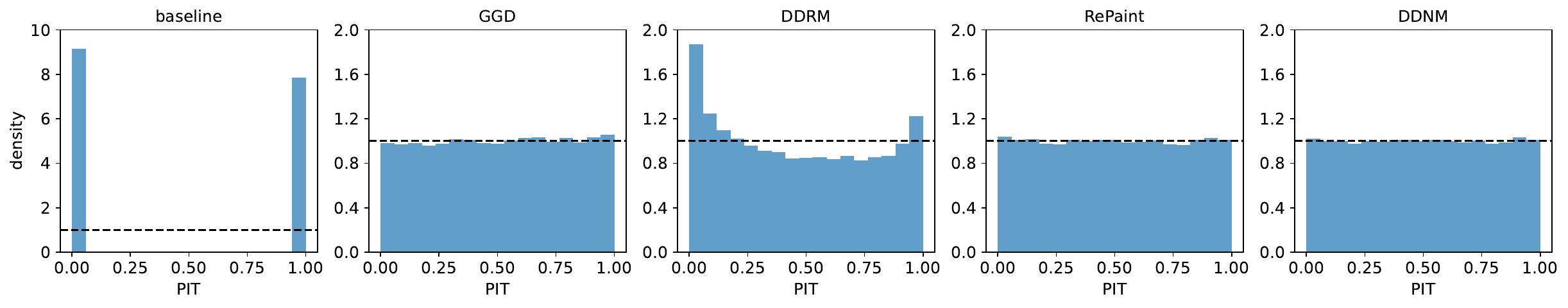}
  \caption{Comparison of PIT distributions for different inpainting algorithms.}
  \label{fig:pit}
\end{figure*}

The results show that RePaint and DDNM have negligible mean $z$-scores, indicating little systematic bias. Their $z$-score widths are also closest to unity. In particular, the DDNM result is consistent with the expected unit width within approximately $1.5\sigma$, indicating the best overall uncertainty calibration among the methods considered. GGD shows a small positive bias and a broader $z$-score distribution, while DDRM exhibits a significant negative bias. 

While the $z$-score evaluates whether the predicted uncertainties are statistically consistent with the true observations, it does not indicate how concentrated or informative the predictions are. Posterior sharpness provides a complementary measure of the concentration of the predictive distribution. Sharpness is calculated as the average pixel-by-pixel posterior standard deviation over the posterior samples for each of the $N$ events. We quantify the per event local posterior spread by
\begin{equation}
\widehat\sigma_{i,p}^2
=
\frac{1}{J-1}
\sum_{j=1}^{J}
\left(
x_{i,p}^{(j)}-\widehat m_{i,p}
\right)^2~,
\end{equation}
where $\widehat m_{i,p}$ was defined in Eq.~\eqref{eq:m_hat}. The sharpness is defined as the mean posterior standard deviation over all
images and pixels in the masked region,
\begin{equation}
S
=
\frac{1}{N|\mathcal D|}
\sum_{i=1}^{N}
\sum_{p\in\mathcal D}
\widehat\sigma_{i,p}.
\end{equation}

The resulting values are also given in Table~\ref{tab:modelzsharpness}. Smaller sharpness corresponds to a more concentrated predictive distribution, but does not by itself indicate better performance, since an overly narrow distribution can be poorly calibrated.

The sharpness values show that DDRM produces the narrowest predictive distribution, while GGD, RePaint, and DDNM have similar sharpness. Considering sharpness together with the $z-$score provides a more complete assessment of the probabilistic reconstruction. RePaint and DDNM combine near-zero mean $z$-scores with widths close to unity, while maintaining similar sharpness. DDNM provides the closest overall agreement with the expected calibration, whereas DDRM, despite having the smallest sharpness, exhibits substantial bias and a $z$-score width significantly above unity. 

The Probability Integral Transform (PIT) provides a complementary, model-independent test of posterior calibration by measuring the percentile of the true calorimeter energy within the predicted distribution. It is computed as follows. For each image $i$ and masked pixel $p\in\mathcal D$, we define the empirical
posterior cumulative distribution function
\begin{equation}
\widehat F_{i,p}(x)
=
\frac{1}{J}
\sum_{j=1}^{J}
\mathbf{1}\!\left(x_{i,p}^{(j)} \le x\right).
\end{equation}
Here $\mathbf{1}(\cdot)$ is the indicator function, equal to one when the condition in the bracket is satisfied and zero otherwise. The corresponding Probability Integral Transform (PIT) value is obtained by
evaluating this empirical CDF at the true masked-pixel value,
\begin{equation}
U_{i,p}
=
\widehat F_{i,p}\!\left(x_{i,p}^{\mathrm{true}}\right)
=
\frac{1}{J}
\sum_{j=1}^{J}
\mathbf{1}\!\left(x_{i,p}^{(j)} \le x_{i,p}^{\mathrm{true}}\right).
\end{equation}
The PIT histogram is then formed from the pooled set of values
\begin{equation}
\left\{U_{i,p}\,:\, i=1,\ldots,N,\; p\in\mathcal D\right\}.
\end{equation}
The PIT therefore represents the fraction of posterior samples that are smaller than the true value. For a calibrated posterior predictive distribution, the PIT values should be
approximately uniformly distributed over the range $[0,1]$.

Figure~\ref{fig:pit} compares the PIT distributions obtained for the different reconstruction algorithms. The baseline method exhibits a pronounced accumulation of events near the boundaries at PIT$=0$ and PIT$=1$, indicating that the true energies frequently lie outside the predicted posterior distribution and that the uncertainty estimates are poorly calibrated. DDRM shows a milder U-shaped distribution with an excess of entries near the boundaries, suggesting that its posterior uncertainties remain somewhat underestimated. In contrast, the GGD, RePaint, and DDNM algorithms produce distributions that are nearly uniform over the full PIT range, demonstrating that their posterior distributions provide relatively accurate statistical description of the true calorimeter energies.

%%%%%%%%%%%%%%%%%%%%%%%%%%%%%%%%%%%%%%%%%%%%%%%%%%%%%%%%%%%%%%
%%%%%%%%%%%%%%%%%%%%%%%%%%%%%%%%%%%%%%%%%%%%%%%%%%%%%%%%%%%%%%

\subsubsection{Empirical coverage}
\label{sec:results:inpainting-comparison:dist}

Empirical coverage provides a direct assessment of the statistical calibration of the predicted posterior uncertainties by comparing the observed coverage of the true calorimeter energies with the nominal confidence levels of the posterior distribution. For a well-calibrated probabilistic model, the empirical coverage should follow the ideal one-to-one relation. For example, a $90\%$ credible interval should contain the true value for approximately $90\%$ of the reconstructed pixels. 

Define the total energy in the dead region $\mathcal{D}$ for the truth image and for each posterior sample as
\begin{equation}
\mathcal{E}_i^{\mathrm{true}}
=
\sum_{p\in\mathcal D} x_{i,p}^{\mathrm{true}},
\qquad
\mathcal{E}_i^{(j)}
=
\sum_{p\in\mathcal D} x_{i,p}^{(j)},
\end{equation}
Let
$\widehat q_i(\alpha)$ denote the empirical $\alpha$-quantile of the set $\{\mathcal{E}_i^{(j)}\}_{j=1}^{J}$. The central credible interval at a given nominal level $q$ is
\begin{equation}
\mathcal I_i(q)
=
\left[
\widehat q_i \left(\frac{1-q}{2}\right),~
\widehat q_i \left(\frac{1+q}{2}\right)
\right].
\end{equation}
The corresponding empirical coverage $C(q)$ at level $q$ is defined as
\begin{equation}
C(q)
=
\frac{1}{N}
\sum_{i=1}^{N}
\mathbf{1}
\left(
\mathcal{E}_i^{\mathrm{true}}
\in
\mathcal I_i(q)
\right).
\end{equation}

\begin{figure}[ht]
\centering
  \includegraphics[width=0.9\columnwidth]{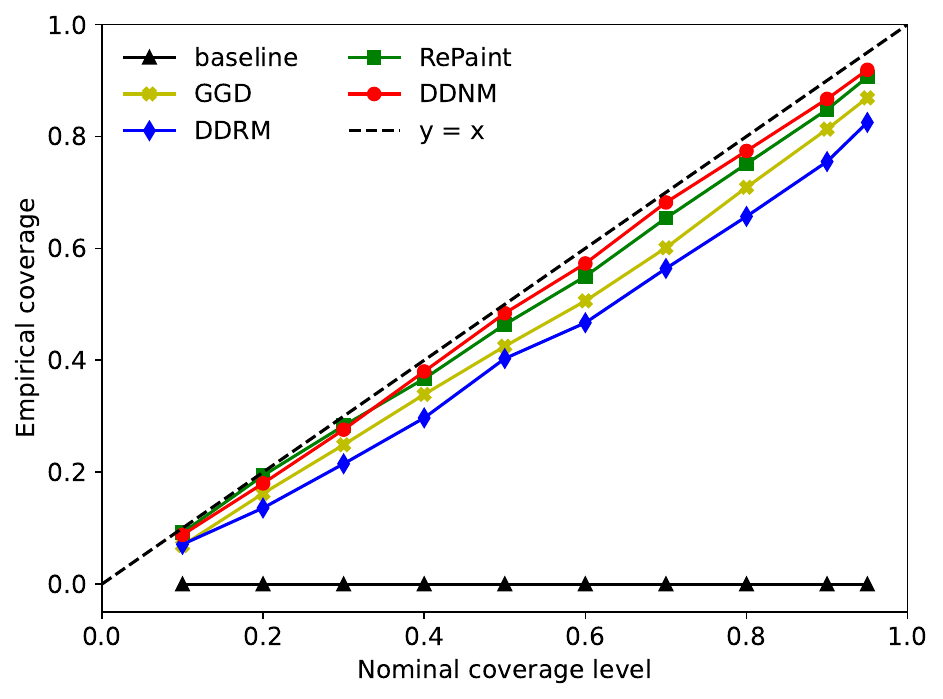}
  \caption{Empirical coverage curves for the different reconstruction methods.}
  \label{fig:coverage}
\end{figure}

Figure~\ref{fig:coverage} compares the empirical coverage curves for the different reconstruction methods. Among these, DDNM and RePaint exhibit the best agreement with the ideal calibration curve over the full range of confidence levels, indicating that their predicted uncertainties capture the statistical variability of the reconstructed energies well. GGD performs similarly with a small systematic deviation from the ideal relation, particularly at higher coverage levels. In contrast, DDRM consistently underestimates the empirical coverage, suggesting that its posterior distributions are overconfident and do not fully capture the uncertainty in the reconstructed energies. The baseline method as expected is substantially worse, with empirical coverage identically zero for all nominal confidence levels.

Overall, the diffusion-based reconstruction methods provide substantially improved performance over the baseline across the deterministic and probabilistic metrics considered. Among the methods evaluated, DDNM provides the most consistent overall performance, combining accurate energy reconstruction, small spatial biases, well-calibrated predictive uncertainties, and a reverse-shrinkage slope consistent with unity. RePaint shows comparable performance, particularly in the $z$-score, PIT, and coverage tests, while GGD also provides reasonable reconstruction quality and uncertainty calibration. DDRM exhibits competitive reconstruction accuracy but poorer uncertainty calibration, despite its relatively small predictive sharpness\footnote{DDRM can be further tuned through its variance-control hyperparameter, $\eta_a$. The default value used in these studies is 0.85; results for a range of values are discussed in Appendix~\ref{app:inverse:ddrm}.}. These results demonstrate that reliable probabilistic reconstruction requires assessing both reconstruction fidelity and uncertainty calibration, with DDNM providing the most balanced performance among the methods studied.

%%%%%%%%%%%%%%%%%%%%%%%%%%%%%%%%%%%%%%%%%%%%%%%%%%%%%%%%%%%%%%
%%%%%%%%%%%%%%%%%%%%%%%%%%%%%%%%%%%%%%%%%%%%%%%%%%%%%%%%%%%%%%

\subsection{Centrality dependence of the reconstruction performance}
\label{sec:results:centrality-comparison}

To investigate the influence of event multiplicity and underlying event activity on the reconstruction quality, the DDNM inpainting algorithm is evaluated for two representative centrality classes corresponding to high- and low-multiplicity Au+Au collisions using a fixed $8\times8$ masked region. These two event classes differ significantly in particle multiplicity and underlying event activity, providing a stringent test of whether the reconstruction performance depends on the complexity of the calorimeter image.
\begin{figure}[ht]
\centering
  \includegraphics[width=\columnwidth]{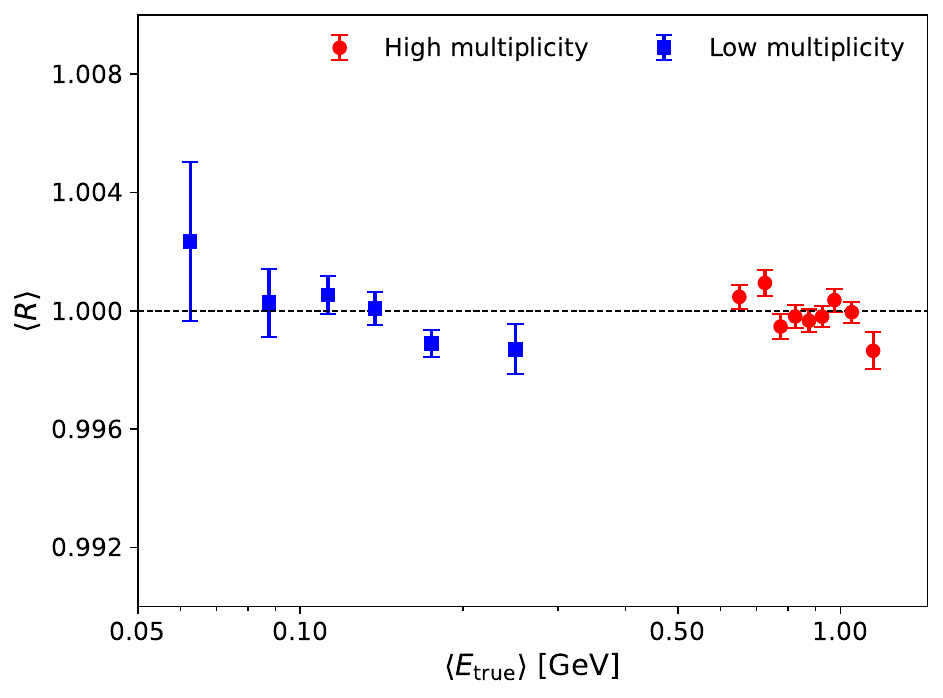}
  \caption{The average ratio of the total reconstructed to total true energy as a function of average true energy for low and high multiplicity events.}
  \label{fig:fullratio_cent}
\end{figure}
Figure~\ref{fig:fullratio_cent} compares the ratio of reconstructed to true full-image energy for the two multiplicity classes. In both cases, the reconstructed energy remains centered close to unity over the full range of average true energies, demonstrating that the inpainting procedure preserves the global event energy without introducing a significant centrality-dependent bias.
\begin{figure}[ht]
\centering
  \includegraphics[width=\columnwidth]{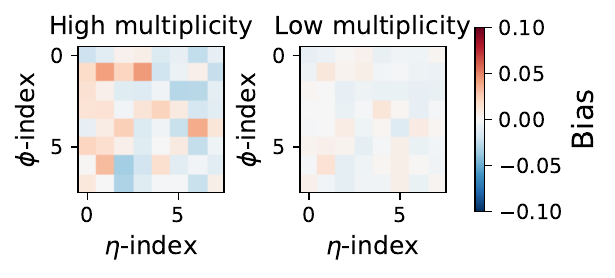}
  \caption{Average residual as a function of $\eta$- and $\phi-$index for low- and high-multiplicity events.}
  \label{fig:bias_cent}
\end{figure}
Figure~\ref{fig:bias_cent} presents the average residual maps for the two centrality classes. In both cases, the residuals fluctuate around zero without exhibiting localized detector-dependent structures, indicating that the reconstruction does not introduce significant position-dependent biases. Although the high-multiplicity events display slightly larger residual fluctuations, their overall magnitude remains small compared to the typical calorimeter energy scale, confirming that the reconstruction remains spatially unbiased even in high-occupancy events.
\begin{table}[ht]
  \centering
  \setlength{\tabcolsep}{4pt}
  \begin{ruledtabular}
  \begin{tabular}{@{}lccc@{}}
    Multiplicity & $\langle z_\mathrm{score}\rangle$ & $\sigma_{z_\mathrm{score}}$ & sharpness \\
    \colrule
        High & 0.01$\pm$0.03 & 1.03$\pm$0.02 & 0.497$\pm$0.003 \\
        Low & 0.03$\pm$0.03 & 1.06$\pm$0.02 & 0.177$\pm$0.002 \\
   \end{tabular}
  \end{ruledtabular}
    \caption{Centrality dependence of $z-$score and sharpness}
    \label{tab:cent_z_sharpness}
\end{table}
The values for $z$-score and sharpness are summarized in Table~\ref{tab:cent_z_sharpness}. For both multiplicity classes, the $z$-score distributions are centered near zero with widths close to unity, indicating good agreement between the predicted uncertainties and the observed fluctuations of the true calorimeter energies. The high-multiplicity events have a larger sharpness than the low-multiplicity events, reflecting the larger uncertainty associated with the more complex collision environment. Importantly, the corresponding $z$-score widths remain close to unity in both cases, indicating that this increase in uncertainty is consistent with the observed event-to-event fluctuations.
\begin{figure}[ht]
\centering
  \includegraphics[width=\columnwidth]{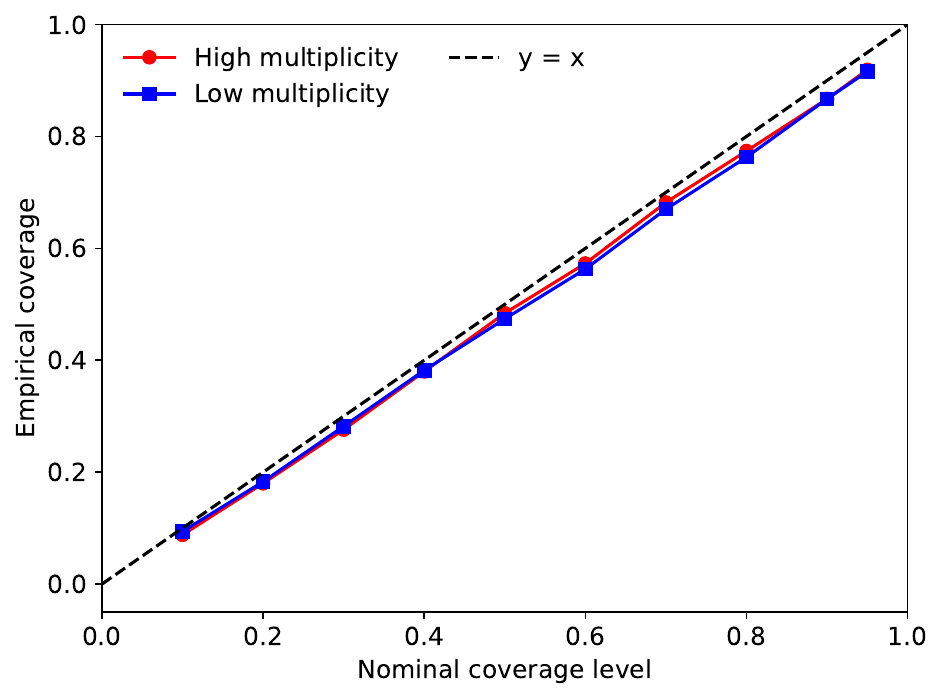}
  \caption{Empirical coverage curves for low- and high-multiplicity event classes.}
  \label{fig:coverage_cent}
\end{figure}
Finally, the empirical coverage curves presented in Figure \ref{fig:coverage_cent} shows close agreement for both multiplicity classes. This agreement indicates that the predicted credible intervals capture the observed statistical fluctuations over a wide range of confidence levels.

Overall, the consistency observed across all reconstruction metrics demonstrates that DDNM-inpainting is largely insensitive to the underlying collision centrality. This robustness is particularly important for heavy-ion analyses, where detector occupancies and background conditions vary substantially with centrality.

%%%%%%%%%%%%%%%%%%%%%%%%%%%%%%%%%%%%%%%%%%%%%%%%%%%%%%%%%%%%%%
%%%%%%%%%%%%%%%%%%%%%%%%%%%%%%%%%%%%%%%%%%%%%%%%%%%%%%%%%%%%%%

\subsection{Dependence on the size of the dead-region}
\label{sec:results:boxsweep}

To investigate the robustness of the proposed reconstruction framework against increasing detector inefficiency, the DDNM algorithm is evaluated for a series of masked regions with progressively increasing box sizes. As the masked region becomes larger, the reconstruction problem becomes increasingly challenging because less surrounding information is available to constrain the posterior prediction.
\begin{figure}[ht]
\centering
  \includegraphics[width=\columnwidth]{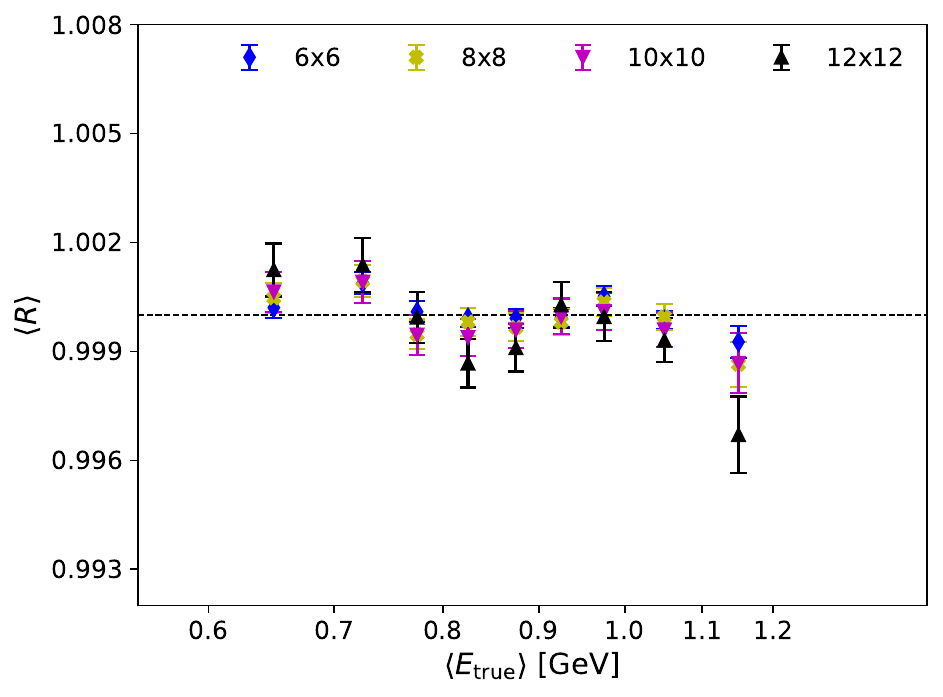}
  \caption{The average ratio of the total reconstructed to total true energy as a function of average true energy for different sizes of the masked region.}
  \label{fig:boxsize_fullratio}
\end{figure}
Figure~\ref{fig:boxsize_fullratio} shows the ratio of the reconstructed and true full-event energies for the different dead-region sizes. The distribution remains centered close to unity over the entire box-size range, demonstrating that the generated samples preserve the total calorimeter energy even when increasingly larger detector regions are removed. Although the event-by-event spread gradually increases with box size, reflecting the larger uncertainty associated with reconstructing more missing information, no significant systematic bias is observed. 
\begin{figure*}[t]
\centering
\includegraphics[width=\textwidth]{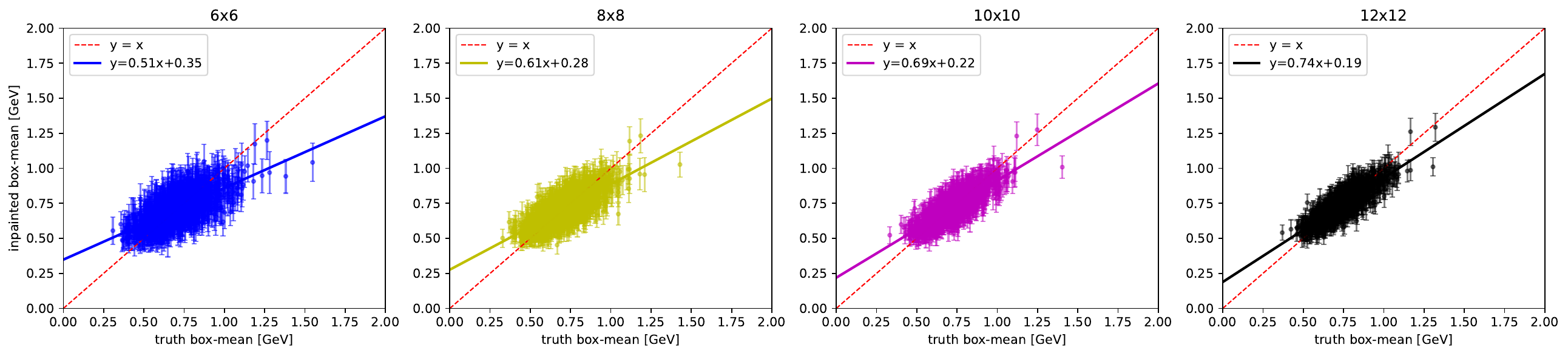}
  \caption{Forward shrinkage of mean reconstructed energy towards true missing energy for different sizes of the dead region. The slope and intercept of the fit are also included.}
  \label{fig:boxsize_shrinkage}
\end{figure*}

\begin{table*}[ht]
  \centering
  \setlength{\tabcolsep}{4pt}
  \begin{ruledtabular}
  \begin{tabular}{@{}lccccc@{}}
   & \multicolumn{2}{c}{Mean}
  & & \multicolumn{2}{c}{Intercept}\\
\cline{2-3} \cline{5-6}
    Box-size & True & Masked & Slope & Expected & Obtained \\
    \colrule
    $6\times6$ & 0.707$\pm$0.002 & 0.710$\pm$0.012 & 0.51$\pm$0.01 & 0.34$\pm$0.01 & 0.35$\pm$0.01\\    
    $8\times8$ & 0.712$\pm$0.002 & 0.711$\pm$0.009 & 0.61$\pm$0.02 & 0.28$\pm$0.01 & 0.28$\pm$0.01 \\
    $10\times10$ & 0.723$\pm$0.002 & 0.721$\pm$0.007 & 0.69$\pm$0.01 & 0.22$\pm$0.01 & 0.22$\pm$0.01 \\
    $12\times12$ & 0.741$\pm$0.001 & 0.738$\pm$0.006 & 0.74$\pm$0.01 & 0.19$\pm$0.01 & 0.19$\pm$0.01 \\
   \end{tabular}
  \end{ruledtabular}
    \caption{Parameters of the forward shrinkage study for different masked box-sizes. The expected intercept is given by $(1- \mathrm{slope})*\mathrm{true \, mean}$.}
    \label{tab:box_size_shrinkage_forward}
\end{table*}
The posterior forward shrinkage results are summarized in Table~\ref{tab:box_size_shrinkage_forward} and shown in Figure~\ref{fig:boxsize_shrinkage}. As the size of the dead region increases, the forward regression slope monotonically increases towards one. This is because of the fact that ${\rm Var}(\varepsilon)$ decreases with box size. This can be seen analytically as follows. From the law of total variance we have ${\rm Var}(\varepsilon) = \mathbb{E}\left({\rm Var}\left[\varepsilon \mid \bfx_{{\rm known}}\right]\right) + {\rm Var}\left(\mathbb{E}\left[\varepsilon \mid \bfx_{{\rm known}}\right]\right)$. The second term is zero due to Eq.~\eqref{eq:T-m-fluctuations}. In the first term, we examine the variance: ${\rm Var}\left[\varepsilon \mid \bfx_{{\rm known}}\right] = \mathbb{E}[\varepsilon^2 \mid \bfx_{{\rm known}}] = {\rm Var}\left[T \mid \bfx_{{\rm known}}\right]$. Since $T=\frac{1}{n^2}\sum_{j=1}^{n^2} x_j$ with $x_j$ being the posterior pixel values of the $n\times n$ masked patch, we have that 
\begin{align}
\label{eq:patchvar}
    {\rm Var}\left(T \mid \bfx_{{\rm known}}\right)= \frac{1}{n^4} \sum_{j,k=1}^{n^2} {\rm Cov}(x_j, x_k \mid \bfx_{{\rm known}})~.
\end{align}
If the per-pixel posterior fluctuations were independent with variance $v$, then ${\rm Var}(\varepsilon)$ collapses to $v/n^2$ (averaging more towers suppresses the fluctuations like one over the number of towers). In calorimeter data however, pixels are correlated, so the suppression is slower than $1/n^2$. However the direction is robust - the larger patch has the smaller ${\rm Var}(\varepsilon)$. Furthermore, in Table \ref{tab:box_size_shrinkage_forward} the observed intercepts agree closely with the expected values for all box sizes.

The results for $z$-score and sharpness with increasing box size are summarized in Table~\ref{tab:size_z_sharpness}. While the $z$-score distributions remain centered around zero, with unit standard deviation, for all dead-region sizes, implying that the reconstructed energies remain largely unbiased, posterior sharpness become slightly broader as the box size increases, reflecting that larger missing regions require more challenging reconstruction task.

\begin{table}[ht]
  \centering
  \setlength{\tabcolsep}{4pt}
  \begin{ruledtabular}
  \begin{tabular}{@{}lccc@{}}
    Box-size & $\langle z_\mathrm{score}\rangle$ & $\sigma_{z_\mathrm{score}}$ & sharpness \\
    \colrule
    $6\times6$ & -0.03$\pm$0.03 & 1.02$\pm$0.02 & 0.495$\pm$0.003 \\
    $8\times8$ & 0.01$\pm$0.03 & 1.03$\pm$0.02 & 0.497$\pm$0.003 \\
    $10\times10$ & 0.03$\pm$0.03 & 0.99$\pm$0.02 & 0.502$\pm$0.003 \\
    $12\times12$ & 0.03$\pm$0.03 & 0.97$\pm$0.02 & 0.514$\pm$0.003 \\
   \end{tabular}
  \end{ruledtabular}
    \caption{Box-size dependence of $z-$score and sharpness}
    \label{tab:size_z_sharpness}
\end{table}

Although larger dead regions lead to broader predictive distributions, the global energy response and uncertainty calibration remain remarkably stable. The agreement of the measured and expected shrinkage intercepts, together with $z$-score widths close to unity across all box sizes, indicates that the reconstruction remains statistically reliable even as the available detector information decreases. This robustness is particularly important for realistic experimental applications, where detector inefficiencies can vary with time and operating conditions, and demonstrates that the proposed approach remains effective across a broad range of detector acceptance scenarios.
\section{Conclusions and Outlook}
\label{sec:conclusion}

In this work, we presented a proof-of-concept study of diffusion-based calorimeter inpainting in a controlled setting, where ground-truth images were generated from the same DDPM that is used as the inpainting prior. The generative model is therefore matched to the data by construction, allowing the performance of different inpainting algorithms to be studied without the additional complication of prior misspecification. This setup provides a controlled assessment of their ability to reproduce the target conditional distribution.

Our results demonstrate that diffusion-based posterior sampling is a viable technique for reconstructing masked regions in calorimeter data. In addition to accurately preserving global calorimeter energy, this procedure results in minimal spatial bias, maintains the largest event-by-event dynamic range, and produces posterior distributions that are well calibrated according to the $z$-score, PIT, and empirical coverage tests. Further, the reconstruction quality also remains stable across the centrality range with no appreciable degradation as the event multiplicity changes. The masked region scan provides an additional demonstration of the robustness. As the masked detector region increases in size, larger predictive uncertainties naturally arise because less information is available to constrain the reconstruction. Nevertheless, the global energy response, spatial uniformity, and posterior calibration remain stable throughout the scan, indicating that the diffusion model adapts its predictive uncertainty appropriately. This is an important property for realistic detector operation, where dead channels and inactive detector regions evolve with time.

Our study demonstrates that the posterior sampling along with the learned diffusion prior is capable of providing meaningful information about unobserved or dead regions in calorimeter data. Several important extensions naturally follow from the present study. Future studies can investigate how probabilistic inpainting influences analysis of physics observables, including jet reconstruction, jet energy resolution, jet substructure, and correlation measurements, where uncertainties in the reconstructed images propagate directly into the final physics results.  Extending the framework to incorporate detector response uncertainties, mixed detector subsystems, and conditional reconstruction strategies represents a promising direction toward a fully probabilistic reconstruction framework for next-generation experiments. 

More broadly, this study provides a controlled validation framework for probabilistic detector reconstruction. Reconstruction bias and energy conservation remain essential, but they do not determine whether an ensemble of reconstructions represents the correct conditional uncertainty. Posterior calibration must therefore be treated as a separate performance criterion. The diagnostics used here are largely model independent and can serve as benchmarks for future probabilistic reconstruction methods beyond diffusion-based approaches.

\begin{acknowledgments}
We thank Axel Drees for useful discussions and Jan C. Bernauer for providing GPU access. The work of RE is supported by US Department of Energy Contract No.~DE-FG02-96ER40988. HR acknowledges support from the Simons Foundation Grant No. 994318.  OpenAI and Anthropic LLMs made significant contributions at various stages of this work. The upstream training data for \cite{torbunov2024calo} was provided by the sPHENIX Collaboration.
\end{acknowledgments}

\appendix
\section{DDPM Details}
\label{app:diffusion}

This appendix collects supplementary material for Sections \ref{sec:setup}, \ref{sec:algorithms} that cover (1) the variance-preserving versus variance-exploding conventions, (2) the forward posterior underlying the reverse step and (3) the Tweedie / score duality behind the denoised estimate $\xhat$.

Notation: $\alpha_t = 1-\beta_t$ is the variance schedule; $\bar\alpha_t = \prod_{i\le t}\alpha_i$ is the cumulative product used for the one-step noising process in Eq.~\eqref{eq:x_t_fast_fwd};
$\net(\bfx_t, t)$ is the trained noise-prediction network; $\bfz_t,\sim\mathcal{N}(\bfzero,\bfI)$ are injected noise at step $t$, and $\xhat$, $\epshat$ are the denoised- and noise-estimates.

\subsection{Variance-preserving and variance-exploding conventions}
\label{app:diffusion:vpve}

Section \ref{sec:setup} adopts the variance-preserving (VP) convention of \cite{ho2020ddpm}, in
which the closed-form marginal of Eq.~\eqref{eq:x_t_fast_fwd} is
\begin{equation}
  q(\bfx_t \mid \bfx_0)
  = \mathcal{N}\!\bigl(\bfx_t;\, \sqrt{\bar\alpha_t}\,\bfx_0,
    \;(1-\bar\alpha_t)\,\bfI\bigr)~,
  \label{eq:app-vp}
\end{equation}
so that the data $\bfx_0$ is progressively attenuated by $\sqrt{\bar\alpha_t}$ while the total variance $\bar\alpha_t + (1-\bar\alpha_t)$ is held fixed to $1$.  Part of the literature on
diffusion-based inverse problems, in particular
DDRM~\cite{kawar2022ddrm}, instead states results in the
variance-exploding (VE) convention, in which the clean signal is left unscaled and noise is added with a growing variance $\sigma_t^2$,
\begin{equation}
  q(\tilde{\bfx}_t \mid \bfx_0)
  = \mathcal{N}\!\bigl(\tilde{\bfx}_t;\, \bfx_0,\; \sigma_t^2\,\bfI\bigr).
  \label{eq:app-ve}
\end{equation}
The two are related by a rescaling factor 
$\tilde{\bfx}_t = \bfx_t/\sqrt{\bar\alpha_t}$, that maps
Eq.~\eqref{eq:app-vp} onto Eq.~\eqref{eq:app-ve} with the variance $\sigma_t^2$ 
\begin{equation}
  \sigma_t^2 = \frac{1-\bar\alpha_t}{\bar\alpha_t}.
  \label{eq:app-vpve-map}
\end{equation}
At $t\to T$, $\bar\alpha_t\to 0$ and $\sigma_t^2$ explodes (hence the name). Because the map is invertible every VE expression can be converted to a VP form. This was used to translate the DDRM update from Ref. \cite{kawar2022ddrm} in Eqns. \eqref{eq:ddrm-dead} and \eqref{eq:ddrm-obs}.

\subsection{Posterior distribution}
\label{app:diffusion:posterior}

In Section \ref{sec:DDPM-setup}, we saw that the reverse conditional $q(\mathbf{x}_{t-1} | \mathbf{x}_{t})$ is intractable. However if we condition on $\bfx_0$, then the conditional $q(\mathbf{x}_{t-1} | \mathbf{x}_{t}, \bfx_0)$ becomes tractable. This is apparent once we rewrite $q(\mathbf{x}_{t-1} | \mathbf{x}_{t}, \bfx_0)$ using Bayes' rule as follows
\begin{align}
    q(\mathbf{x}_{t-1} \mid \mathbf{x}_{t}, \bfx_0) = \frac{q(\mathbf{x}_{t} \mid \mathbf{x}_{t-1}, \bfx_0)q(\mathbf{x}_{t-1} \mid \bfx_0)}{q(\mathbf{x}_{t} \mid \bfx_0)}~.
\end{align}
Since the forward noising process is Markov we have $q(\mathbf{x}_{t} \mid \mathbf{x}_{t-1}, \bfx_0) = q(\mathbf{x}_{t} \mid \mathbf{x}_{t-1})$. All the conditionals appearing in the right hand side above are Gaussians and explicitly known. It therefore follows that
\begin{align}
  \label{eq:app-posterior}
  q(\bfx_{t-1} \mid \bfx_t, \bfx_0)
  &= \mathcal{N}\!\left(\bfx_{t-1};\,
     \tilde{\bmu}_t(\bfx_t, \bfx_0),\; \tilde\beta_t \bfI\right),
\end{align}
where the posterior mean and variance are
\begin{align}
  \label{eq:app-posterior-mean}
  \tilde{\bmu}_t &= \frac{\sqrt{\bar\alpha_{t-1}}\,\beta_t}{1-\bar\alpha_t}\, \bfx_0
   + \frac{\sqrt{\alpha_t}\,(1-\bar\alpha_{t-1})}{1-\bar\alpha_t}\, \bfx_t~,\\[5pt]
  % \label{eq:app-posterior-varaince}
  \label{eq:beta-tilde}
  \tilde\beta_t &= \frac{1-\bar\alpha_{t-1}}{1-\bar\alpha_t}\,\beta_t~.
\end{align}
The reverse DDPM step is born out of this conditional
\begin{align}
  \bfx_{t-1}&=\ddpmstep(\bfx_{t}, \bfx_0) = \nonumber\\[5pt]
  &\frac{\sqrt{\bar\alpha_{t-1}}\,\beta_t}{1-\bar\alpha_t}\;\bfx_0
   + \frac{\sqrt{\alpha_t}\,(1-\bar\alpha_{t-1})}{(1-\bar\alpha_t)}\; \bfx_t + \sqrt{\tilde{\beta}_t}\;\zeta_t~,
  \label{eq:ddpmstep}
\end{align}
However at sampling time, we do not have the information about the clean state $\bfx_0$ so the above equation as written is not practical. To circumvent this, a one step MSE optimal estimate $\mathbb{E}[\bfx_0 \mid \bfx_t]$ based on the current noisy state $\bfx_t$ is obtained via Tweedie'{}s formula~\cite{efron2011tweedie} in terms of the marginal score $\nabla_{\bfx_t} \log q_t(\bfx_t)$
\begin{equation}
  \mathbb{E}[\bfx_0 \mid \bfx_t]
  = \frac{\bfx_t + (1-\bar\alpha_t)\,\nabla_{\bfx_t} \log q_t(\bfx_t)}
         {\sqrt{\bar\alpha_t}} ~.
  \label{eq:app-tweedie}
\end{equation}
The score function can be expressed as a conditional expectation of the forward noise $\epsilon$\footnote{Eq. \eqref{eq:app:score-exp} follows from the Fisher identity $\nabla_{\bfx_t} \log q_t\left(\bfx_t\right)=\mathbb{E}_{\bfx_0 \sim q(\bfx_0 \mid \bfx_t)}\left[\nabla_{\bfx_t} \log q\left(\bfx_t \mid \bfx_0\right) \mid \bfx_t\right]$ and the fact that $\bfx_t \mid \bfx_0$ is gaussian: $x_t=\sqrt{\bar{\alpha}_t} x_0+\sqrt{1-\bar{\alpha}_t} \epsilon, \quad \epsilon \sim \mathcal{N}(0, I)$.}
\begin{align}\label{eq:app:score-exp}
\nabla_{\bfx_t} \log q_t\left(\bfx_t\right)=-\frac{\mathbb{E}\left[\epsilon \mid \bfx_t, t\right]}{\sqrt{1-\bar{\alpha}_t}} .
\end{align}
A noise-prediction network trained with the standard mean-squared-error objective in Eq.~\eqref{eq:DDPM_objective} estimates the conditional expectation,
\begin{align}
\epsilon_\theta\left(\bfx_t, t\right) \approx \mathbb{E}\left[\epsilon \mid \bfx_t\right] .
\end{align}
Substituting this estimate into Tweedie's formula in Eq.~\eqref{eq:app-tweedie} yields the one-step denoised estimate
\begin{align}
  \label{eq:tweedie}
  \xhat(\bfx_t)
  = \frac{\bfx_t - \sqrt{1-\bar\alpha_t}\;\net(\bfx_t, t)}
         {\sqrt{\bar\alpha_t}}~.
\end{align}
Replacing $\bfx_0$ with $\xhat$ in the DDPM reverse step Eq.~\eqref{eq:ddpmstep} finally yields the expression in Eq.~\eqref{eq:ddpm-step} one uses for sampling. 

To summarize, we start by sampling $\bfx_T\sim \mathcal{N}(\bfzero, \bfI)$, iterating down the DDPM steps Eq.~\eqref{eq:ddpm-step} from $t= T, \cdots, 2$. In the final step $t=1$ the noise variance vanishes $\tilde{\beta}_1 = 0$ (since $\bar{\alpha}_0=1$) and we get the final denoised estimate $\bfx_0 = \xhat(\bfx_1)$ from Eq.~\eqref{eq:tweedie}. 

DDIM~\cite{song2021ddim} generalizes the reverse update to a non-Markovian family that preserves
the same forward marginals
\begin{equation}
\label{eq:ddimstep}
\bfx_{t-1}
=
\sqrt{\bar{\alpha}_{t-1}}\,
\hat{\bfx}_0(\bfx_t)
+
\sqrt{1-\bar{\alpha}_{t-1}-\sigma_t^2}\,
\epsilon_\theta(\bfx_t,t)
+
\sigma_t \zeta_t~,
\end{equation}
where $\zeta_t\sim\mathcal{N}(\bfzero,\mathbf I)$ and $0\leq \sigma_t\leq\sqrt{1-\bar{\alpha}_{t-1}}$. The choice $\sigma_t=0$ gives the deterministic DDIM update, whereas $\sigma_t^2=\widetilde{\beta}_t$ reduces it to the DDPM transition. The DDRM dead-pixel update in Eq.~\eqref{eq:ddrm-dead}, has the same structural form but chooses the variance $\sigma_t=\eta_a\sqrt{1-\bar{\alpha}_{t-1}}$.  To match it to the DDPM transition would require $\eta_a$ to be $t$ dependent
\begin{equation}
\eta_a^2=\frac{\widetilde{\beta}_t}{1-\bar{\alpha}_{t-1}}~.
\end{equation}
Consequently, a constant $\eta_a$ cannot reproduce the full DDPM transition.
\section{Optimal denoiser}
\label{app:inverse:optimal-denoiser}

Let $\bfx_0\sim p_{\rm data}$ with (finite) mean $\mu = \mathbb{E}[x_0]$ and covariance $\Sigma = \operatorname{Cov}(x_0)$. The optimal denoiser is the following minimum-mean-square-error estimator
\begin{align}
\label{eq:D-def-1}
  D(\bfx_t) \;:=\; \mathbb{E}\!\left[\bfx_0 \,\middle|\, \bfx_t\right].
\end{align}
Since $\bfx_t = \sqrt{\bar{\alpha}_t} \bfx_0 + \sqrt{1-\bar{\alpha}_t} \zeta$ with $\zeta \sim \mathcal{N}(\bfzero, \bfI)$, it follows that the marginal density of $\bfx_t$ is
\begin{align}
\label{eq:pt-marginal}
    p_t(\bfx) \;=\; \int d\bfx_0~p_{\rm data}(\bfx_0)\,\mathcal{N}(\bfx;\,\sqrt{\bar{\alpha}_t} \bfx_0,\,(1-\bar{\alpha}_t) \bfI).
\end{align}
This allows to write $D(\bfx)$ more explicitly as 
\begin{align}
\label{eq:D-def-2}
    D(\bfx) \;=\; \frac{1}{p_t(\bfx)}\int d\bfx_0~\bfx_0p_{\rm data}(\bfx_0)\,\mathcal{N}(\bfx;\,\sqrt{\bar{\alpha}_t} \bfx_0,\,(1-\bar{\alpha}_t) \bfI)~.
\end{align}
Differentiating the expression of $p_t(\bfx)$ above w.r.t. $\bfx_t$ yields
\begin{align}
  \nabla_{\bfx} p_t(\bfx)= \frac{p_t(\bfx)}{1-\bar{\alpha}_t}\bigl(\sqrt{\bar{\alpha}_t}\,D(\bfx) - \bfx\bigr),
\end{align}

We note that rearranging the above, the optimal denoiser depends on the score function $\nabla_{\bfx} \log p_t(\bfx)$ as follows
\begin{align}
\label{eq:optimal-denoiser-2}
  D(\bfx) = \frac{\bfx+(1-\bar{\alpha}_t)\nabla_{\bfx} \log p_t(\bfx)}{\sqrt{\bar{\alpha}_t}}
\end{align}

\begin{figure*}[!htb]
  \includegraphics[width=\textwidth]{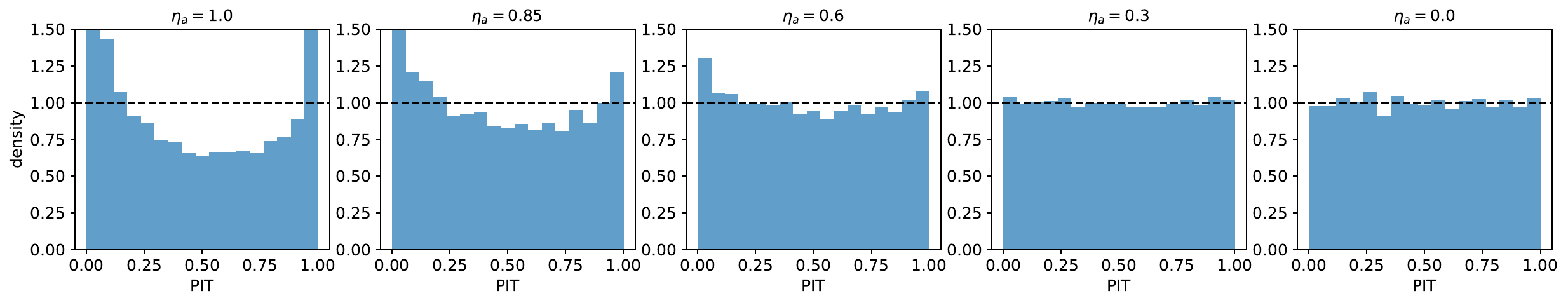}
  \caption{Comparison of PIT histograms for DDRM fills for $\eta_a\in\{0,0.3,0.6,0.85,1\}$}
  \label{fig:ddrm-eta-pit}
  \includegraphics[width=\textwidth]{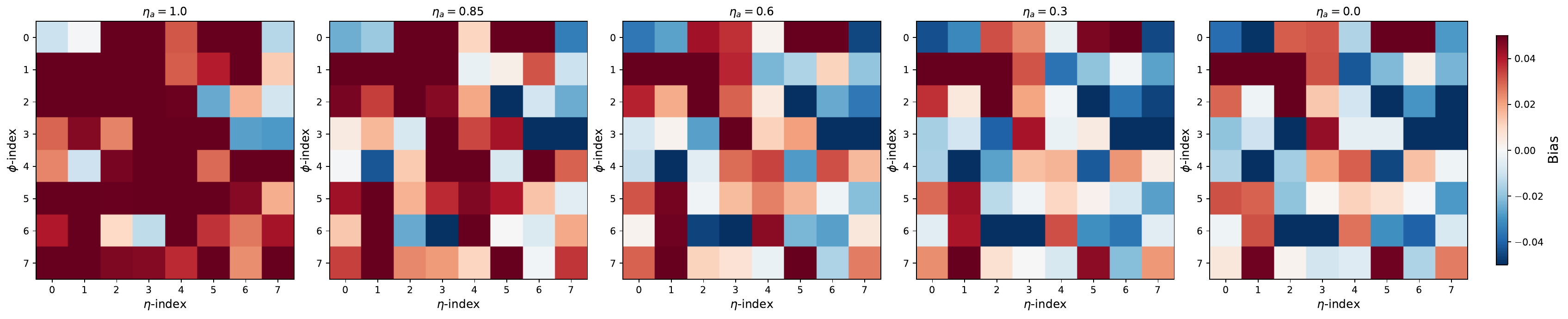}
  \caption{Comparison of mean spatial bias for DDRM fills for $\eta_a\in\{0,0.3,0.6,0.85,1\}$.}
  \label{fig:ddrm-eta-mean-bias}
\end{figure*}

A further differentiation gives
\begin{align}
  \label{eq:hessian}
  \nabla_{\bfx}^2 \log p_t(\bfx)
  = \frac{\bar{\alpha}_t}{\left(1-\bar{\alpha}_t\right)^2}\operatorname{Cov}\!\left[\bfx_0 \mid \bfx_t = \bfx\right]
    \;-\; \frac{1}{1-\bar{\alpha}_t}\,\bfI~.
\end{align}
Applying $\nabla_{\bfx}$ to Eq.~\eqref{eq:optimal-denoiser-2} and substituting $\nabla_{\bfx}^2 \log p_t(\bfx)$ from Eq.~\eqref{eq:hessian}, we find the following result:
\begin{align}
\label{eq:jacobian}
    \nabla_{\bfx}D(\bfx) &= \frac{1}{\sqrt{\bar{\alpha}_t}}\Bigl(\bfI + (1-\bar{\alpha}_t)\,\nabla_\bfx^2 \log p_t(\bfx)\Bigr)~,\nonumber\\[5pt]
    &=\frac{\sqrt{\bar{\alpha}_t}}{1-\bar{\alpha}_t}\operatorname{Cov}\!\left[\bfx_0 \mid \bfx_t = \bfx\right]
\end{align}
This result exposes the fact that the local sensitivity of the optimal denoiser is the posterior uncertainty rescaled by a factor of $\sqrt{\bar{\alpha}_t} / \left(1-\bar{\alpha}_t\right)$

In the limit $t\to T$ we have $\sqrt{\bar{\alpha}_t}\to 0$. As a result the posterior distribution $p(\bfx_0 \mid \bfx_t)$
\begin{align}
    &p(\bfx_0 \mid \bfx_t) \propto\nonumber\\ 
    &p_{\rm data}(\bfx_0) \exp\left( \frac{\sqrt{\bar{\alpha}_t}}{1-\bar{\alpha}_t} \bfx_0 \cdot \bfx - \frac{\bar{\alpha}_t}{2(1-\bar{\alpha}_t)}\|\bfx_0\|^2\right)
\end{align}
reduces to the prior $p_{\rm data}(\bfx_0)$ in the limit. Consequently, $\operatorname{Cov}\!\left[\bfx_0 \mid \bfx_t = \bfx\right] \to \Sigma$, a fixed finite matrix that is independent of $\bfx$. This proves that the gradient of the optimal denoiser is $O(\sqrt{\bar{\alpha}_t})$ in the limit of $t\to T$. 

%%%%%%%%%%%%%%%%%%%%%%%%%%%%%%%%%%%
%%%%%%%%%%%%%%%%%%%%%%%%%%%%%%%%%%%

\section{DDRM $\eta_a$ sweep}
\label{app:inverse:ddrm}

The DDRM Algorithm \ref{alg:ddrm} contains the hyperparameter $\eta_a\in[0,1]$, which governs the stochasticity of the update rule in the dead region in Eq.~\eqref{eq:ddrm-dead}. In the main text, we presented results for the default value $\eta_a=0.85$ and saw signs of miscalibration in the statistical tests in Section \ref{sec:results:inpainting-comparison}. Here we generate DDRM fills, keeping $\eta_b$ fixed to $1$, and sweep $\eta_a\in\{0,0.3,0.6,0.85,1\}$. The result for the PIT histograms and mean spatial bias are shown in Figures \ref{fig:ddrm-eta-pit} and \ref{fig:ddrm-eta-mean-bias} respectively. The value $\eta_a=1$, shows strong U-shaped distribution indicating substantial miscalibration. The deviation from uniformity decreases systematically as $\eta_a$ is reduced, with smaller values of $\eta_a$ giving nearly uniform PIT distributions. Likewise for smaller values of $\eta_a$ the mean spatial bias is more uniform about zero. Smaller values of $\eta_a$ yield substantially better calibrated DDRM samples.

\newpage
\bibliography{references}

\end{document}